\documentclass[fleqn,usenatbib]{mnras}

\usepackage{newtxtext,newtxmath}
\usepackage{supertabular,booktabs}
\usepackage{scalefnt}

\usepackage[T1]{fontenc}

\DeclareRobustCommand{\VAN}[3]{#2}
\let\VANthebibliography\thebibliography
\def\thebibliography{\DeclareRobustCommand{\VAN}[3]{##3}\VANthebibliography}

\usepackage{graphicx}	
\usepackage{amsmath}	
\usepackage{newtxtext,newtxmath}
\usepackage{enumitem}

\newcommand{\orcid}[1]{\href{https://orcid.org/#1}{\includegraphics[width=10pt]{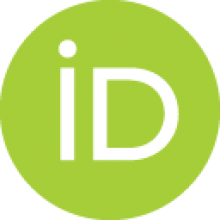}}}

\newcommand{\pp}{$\phantom{-}$}

\newbox\grsign \setbox\grsign=\hbox{$>$} \newdimen\grdimen \grdimen=\ht\grsign
\newbox\simlessbox \newbox\simgreatbox
\setbox\simgreatbox=\hbox{\raise.5ex\hbox{$>$}\llap
     {\lower.5ex\hbox{$\sim$}}}\ht1=\grdimen\dp1=0pt
\setbox\simlessbox=\hbox{\raise.5ex\hbox{$<$}\llap
     {\lower.5ex\hbox{$\sim$}}}\ht2=\grdimen\dp2=0pt

\newbox\simppropto
\setbox\simppropto=\hbox{\raise.5ex\hbox{$\sim$}\llap
     {\lower.5ex\hbox{$\propto$}}}\ht2=\grdimen\dp2=0pt

\title[Reanalysis of neutron-capture elements in the benchmark r-rich star CS 31082-001]{Reanalysis of neutron-capture elements in the benchmark r-rich star CS 31082-001}
\author[H. Ernandes et al.]{
H. Ernandes$^{1,2,3}$\orcid{0000-0001-6541-1933},\thanks{E-mail: heitor.ernandes@astro.lu.se}
M. J. Castro$^{1}$\orcid{0000-0002-3792-8043},
B. Barbuy$^{1}$\orcid{0000-0001-9264-4417},  
M. Spite$^{4}$,
\,V. Hill$^{5}$\orcid{0000-0002-8852-3708},
B. Castilho$^{6}$\orcid{0000-0002-6696-2611},
C. J. Evans$^{7}$
\\
$^{1}$Universidade de São Paulo, IAG, Rua do Matão 1226, Cidade Universitária, São Paulo 05508-900, Brazil\\
$^{2}$Lund Observatory, Department of Astronomy and Theoretical Physics, Lund University, Box 43, SE-22100 Lund, Sweden \\
$^{3}$Lund Observatory, Department of Geology, S\"olvegatan 12, SE-22362 Lund, Sweden \\
$^{4}$GEPI, Observatoire de Paris, PSL Research University, CNRS,
Place Jules Janssen, 92190 Meudon, France\\
$^{5}$Universit\'e de Sophia-Antipolis,
 Observatoire de la C\^ote d'Azur, CNRS UMR 6202, BP4229, 06304 Nice Cedex 4, France\\
$^{6}$Laborat\'orio Nacional de Astrof\'{\i}sica, Minist\'erio da Ci\^encia,
Tecnologia e Inova\c{c}\~oes - LNA/MCTI, R. dos Estados Unidos, 154,  Itajub\'a, 37504-364, Brazil\\
$^{7}$ European Space Agency (ESA), ESA Office, Space Telescope Science Institute, 3700 San Martin Drive, Baltimore, MD 21218, USA}

\date{Accepted XXX. Received YYY; in original form ZZZ}

\pubyear{2023}

\begin{document}
\label{firstpage}
\pagerange{\pageref{firstpage}--\pageref{lastpage}}
\maketitle

\begin{abstract} 
We revisit the abundances of neutron-capture elements in the metal-poor ([Fe/H]\,$=$\,$-$2.9) r-process-rich halo star CS~31082-001.
Partly motivated by the development of the new near-ultraviolet Cassegrain U-band Efficient Spectrograph for the Very Large Telescope, we compiled an expanded line list for heavy elements over the range 3000-4000~{\rm \AA}, including hyperfine structure for several elements. Combining archival near-ultraviolet spectra of CS~31082-001 from the Hubble Space Telescope and the Very Large Telescope, we investigate the abundances and nucleosynthesis of 35 heavy elements (Ge, Sr, Y, Zr, Nb, Mo, Ru, Rh, Pd, Ag, Cd, Sn, Ba, La, Ce, Pr, Nd, Sm, Eu, Gd, Tb, Dy, Ho, Er, Tm, Yb, Lu, Hf, Os, Ir, Pt, Pb, Bi, Th, and U). Our analysis includes the first abundance estimates for tin, holmium, and ytterbium  from these data, and the first for lutetium from ground-based data, enabling a more complete view of the abundance pattern of this important reference star. In general, the r-process dominated elements are as enhanced as those in the Sun, particularly for elements with Z\,$\ge$\,56 (Ba and heavier). However, the abundances for the lighter elements in our sample, from Ge to Sn (31\,$\le$\,Z\,$\le$\,50), do not scale with the solar abundance pattern. Moreover, the Ge abundance is deficient relative to solar, indicating that it is dominantly an iron-peak rather than neutron-capture element. Our results (or upper limits) on Sn, Pt, Au, Pb and Bi all pose further questions, prompting further study on the origin and evolution of the known r-rich and actinide-rich, metal-poor stars.

\end{abstract}

\begin{keywords}
Abundances stars: abundances – stars: atmospheres – stars: individual: BPS CS 31082-001; Galaxy: halo.
\end{keywords}



\section{Introduction}

The detailed study of heavy-element stellar abundances can give us new insights into the nuclear processes that created them. In the wider context of galaxy archaeology, quantitative stellar abundances can also provide a powerful discriminator of the different phenomena contributing to nucleosynthesis. For instance, stars on the asymptotic giant branch (AGB) and spinstars (Cescutti et al. 2013, Frischknecht et al. 2016) are often associated with the synthesis of s-process elements. Meanwhile, there are several candidates for r-process enrichment, including: core-collapse supernovae (SNe), magnetorotationally-driven supernovae (Winteler et al. 2012; M\"osta et al. 2018),  jet-driven supernovae (Fujimoto et al. 2008; Nishimura et al. 2015), neutron-neutron star mergers (Abbott et al. 2017), and black-hole-neutron star mergers (Just et al. 2015).

Here we revisit the neutron-capture elements in the well-studied halo star CS~31082-001, which is known to be an r-process and actinide-rich star. CS~31082-001 has a metallicity of [Fe/H]\,$=$\,$-$2.9\,$\pm$\,0.1 (Hill et al. 2002), a europium enhancement of [Eu/Fe]\,$=$\,$+$1.62 (with [Ba/Eu]\,$=$\,$-$0.59), and a clear enhancement of the actinide element uranium. It is also among the brightest very metal-poor stars, with $V$\,$=$\,11.7\,mag.

Given its abundances, CS~31082-001 is classified as a r-II star, a class which gathers stars with  [Eu/Fe]\,$>$\,$+$1.0 and [Ba/Eu]\,$<$\,0.0 (Beers \& Christlieb 2005). A more recent criterion of the class is a modified value of [Eu/Fe]\,$>$\,$+$0.7 (Holmbeck et al. 2020), and an interesting suggestion  by Roederer et al. (2018) is that stars with [Eu/Fe]\,$>$\,$+$0.7 are only found in halo-like orbits and that they were probably formed in low star-formation efficiency environments such as those found in dwarf galaxies.

 In a recent study, Ernandes et al. (2022, hereafter E22) employed archival near-UV spectroscopy of CS~31082-001 from ground and space observations to revisit its chemical abundances of light and iron-peak elements (Z $<$ 32). This led to the first abundance estimates for Be, V and Cu for the star and enabled new comparisons with nucleosynthesis models.  Here we turn our attention to the heavy elements (Z $>$ 32) in CS~31082-001. 
 
 The near-UV region is particularly rich for the study of heavy elements, making it a critical input to studies of Galactic chemical evolution and investigations of the slow and rapid neutron-capture processes and the astrophysical sites where they occur. To expand on past analyses of the heavy elements in CS~31082-001, we compiled a broader line list than in previous studies. The advantage of building on the previous works is that there are already estimated abundances for many elements, which helps to mitigate the uncertainties due to blends when extending the analysis to new lines or species. With an updated and expanded description of the elemental abundances of CS~31082-001, our objective is to investigate the predicted nucleosynthesis yields from different sources.



We add that although the near-UV region has a key role in studies of stellar nuclear processes, it is a relatively under-explored part of the spectrum. This is primarily due to the increasingly high atmospheric absorption towards shorter wavelengths from the ground, resulting in zero transmission below 3000\,\AA\ (although wavelengths below this can be accessed from space, with e.g. the Hubble Space Telescope, HST). Indeed, part of our motivation for this study is the development of the new Cassegrain U-Band Efficient Spectrograph (CUBES) instrument for the Very Large Telescope (VLT). CUBES will provide unprecedented efficiency over the spectral range of 3000-4050~{\rm \AA} with two spectral resolving powers ($R$\,$\sim$\,7,000 and $>$20,000). Using its higher spectral resolving power, CUBES will be able to obtain near-UV spectroscopy of targets which are two-to-three magnitudes fainter than currently possible, e.g. reaching a signal-to-noise (S/N) of 20 at 3130~{\rm \AA} in 1\,hr of exposure for an A0-type star with $U$\,$=$\,17.5\,mag. (Cristiani et al. 2022). This will bring an exciting new capability for studies of metal-poor stars similar to CS~31082-001.

The archival spectra used in the analysis are summarised briefly in Section~2, with the analysis and previous results discussed in Section~3.  The heavy elements studied are detailed in Section~4, including the relevant calculations and fitting methods, with a discussion of the results in Section~5, and our conclusions in Section~6.  Supporting material is given in the Appendices, including the near-UV line list of the heavy elements studied in CS~31082-001, and simulated CUBES observations of two illustrative heavy-element lines in its spectrum.

\section{Observations}

To investigate the heavy-element lines in CS~31082-001 we used the same near-UV spectra as in E22, which were obtained with the Ultraviolet and Visual Echelle Spectrograph (UVES, Dekker et al. 2000) at the VLT in August and October 2000 in the `First Stars' programme (ID 165.N-0276(A), PI: R. Cayrel). Here we employ the spectra observed with central wavelengths of 3400\,\AA, which covers the 3000-3850~{\rm \AA} range, and the spectra centered at 4370~{\rm \AA}, covering 3770-4990~{\rm \AA} (Hill et al. 2002).  As in E22,  the three spectra observed at the 3400~{\rm \AA} setting were combined, giving a S/N ratio of 100 at 3400~{\rm \AA}, and 20 at 3070 {\rm \AA} (Spite et al. 2005, Barbuy et al. 2011, Siqueira-Mello et al. 2013).

For first estimates, we derived all abundances from a smoothed spectrum (FWHM\,$\approx$\,0.09\,\AA). To then estimate the error in the line fitting and choice of continuum, we fitted all the lines in the raw (un-smoothed) data, which have a FWHM\,$\approx$\,0.06 to 0.08\,\AA. Where possible, the initial fits adopted the previous abundances collated by E22, and for this reason the values can appear very specific; for example, A(Y)\,$=$\,$-$0.23, which when also computed with $\pm$0.2, gives $-$0.03 and $-$0.43.

A further check was carried out using spectra observed at the Keck telescope with the HIRES 
spectrograph (Sneden et al. 2009), where we retrieved and combined the blue orders from the observations that overlap with the wavelength range of our selected UVES data. After summing the observations for each order and normalising them, we combined the orders together for a final spectrum (enabled by the overlap of $\sim$5\,\AA\ between each order and because the HIRES spectra have the same rebinned wavelength sampling as the UVES data).

For lines near the ground atmospheric UV cut-off (i.e. 3000-3070~{\rm \AA}) we also use spectra from 
the HST Space Telescope Imaging Spectrograph (STIS) with the E230M grating (see Barbuy et al. 2011), from which we use the data from order~2 (3012-3070~{\rm \AA}) and order~3 (2980-3025~{\rm \AA}).

Details of the observations analysed in this work are summarised in Table~\ref{keck}, including spectral resolving power ($R$), mean S/N per pixel\footnote{Calculated with http://www.stecf.org/software/ASTROsoft/DER\_SNR/}, wavelength range, program identifiers and Principal Investigator (PI).

\setlength{\tabcolsep}{6pt}
\begin{table}
\scalefont{0.88}
\caption{Summary of observations used in this study. Columns are: the spectral resolving power ($R$), the mean signal-to-noise (S/N) per rebinned pixel, the wavelength range, the program identifiers (ID) and Principal Investigators (PI).}
\label{keck}  
\centering                  
\begin{tabular}{lcccccccc} 
\hline\hline             
Instrument & $R$ & S/N & $\lambda$-range ({\rm \AA}) & ID & PI \\ 
\hline  
VLT-UVES & 42,000 & 150  & 3020-3810 & 165.N-0276 & R. Cayrel \\
VLT-UVES & 47,000 & 350  & 3730-4993 & 165.N-0276 & R. Cayrel \\
Keck-HIRES & 40,000 & 150 & 3201-4718 & U53H & M. Bolte \\
HST-STIS & 30,000 & 40  & 2680-3070 & 9359 & R. Cayrel \\
\hline
\hline       
\end{tabular}
\end{table}

The reduced UVES data analysed here were reprocessed by ESO in their archive in 2020. These reductions include several developments compared to the UVES data analysed in previous papers.  When the {\sc midas} version of the pipeline was ported to the ESO Common Pipeline Library (CPL), an important step was the implementation of a proper optimal extraction. Additional improvements included order tracing, robustness to bad pixels (in particular column traps affecting one of the two chips) and enabling successful extractions even when one order lacks a signal (interpolating the results from traces on adjacent orders), and improved calibration of spectral response and flux (see Larsen et al. 2007).  The pipeline was further upgraded for implementation in the EsoReflex environment (Freudling et al. 2013).

\section{Abundance analysis}

For the abundance analysis we used the {\tt Turbospectrum} code from Alvarez \& Plez (1998) and Plez (2012) to generate the synthetic spectra. For stellar parameters we adopted (T$_{\rm eff}$, log~g, [Fe/H], v${\rm t}$) = (4825\,$\pm$\,50 K, 1.5\,$\pm$\,0.3, $-$2.9\,$\pm$\,0.1, 1.8\,$\pm$\,0.2 km\,s$^{-1}$) from Hill et al. (2002).  Model atmosphere grids are from Gustafsson et al. (2008).


The list of lines of heavy elements, together with oscillator strengths  from 
Kurucz (1993)\footnote{http://kurucz.harvard.edu/atoms.html}, and the Vienna Atomic Line Database 
(VALD; Piskunov et al. 1995; Ryabchicova et al. 2015)\footnote{http://vald.astro.uu.se/}, 
are given in Table \ref{Alllines}. The wavelengths are mostly from Kurucz (1993).

Compared to previous analyses, we highlight the inclusion of \ion{Ho}{II} and \ion{Yb}{II}.  We also added more lines for other elements to extend the line list, and checked these one-by-one if they were too faint or blended to be useful (see comments in Section~4). We were able to include lines for: Ge, Sr, Y, Zr, Nb, Mo, Ru, Rh, Pd, Ag, (Cd\footnote{Unfortunately the single \ion{Cd}{I} line in the near-UV region is too weak to derive an elemental abundance.}), Sn, Ba, La, Ce, Pr, Nd, Sm, Eu, Gd, Tb, Dy, Ho, Er, Tm, Yb, Lu, Hf, Os, Ir, Pt, (Pb\footnote{The two \ion{Pb}{I} lines in this region are not sensitive enough to derive an elemental abundance. Therefore, we adopted the results from Plez et al. (2004) and Siqueira-Mello et al. (2013).}), Bi, Th, and U. A blended line of Sn can be used with uncertainties.

\medskip
We present our estimated abundances for CS~31082-001 from the expanded line list in Table~\ref{heavy}; these include the first estimates of Sn, Ho and Yb from these data, as well as for Ba and Lu from analysis of the ground-UV lines. In the table we also compile previous abundance estimates for the neutron-capture elements from Hill et al. (2002), Plez et al. (2004), Barbuy et al. (2011) and Siqueira-Mello et al. (2013), as given by E22 and Sneden et al. (2009). Solar abundances from Asplund et al. (2021) are also given in the third column of the table. We note that the latter are somewhat different from the solar abundances adopted by Hill et al. (2002), which were from Grevesse \& Sauval (1998).

\subsection{Hyperfine structure: Summary}

Some of the heavy elements that we investigate here present hyperfine structure (HFS), which we now describe briefly in turn:

\begin{itemize}
     \item {\it  Barium:} for the unique \ion{Ba}{II} line available in the ground UV we computed the HFS. \smallskip

    \item {\it  Lanthanum:} the HFS shifts for \ion{La}{II} lines are given by Lawler et al. (2001a), and they are available in the VALD line lists.\smallskip

    \item {\it  Cerium:} according to Lawler et al. (2009), there is no need to include the HFS for \ion{Ce}{II} because the main isotopes are even with nuclear spin I\,$=$\,0.\smallskip

    \item {\it  Praseodymium:} HFS from Sneden et al. (2009) is included in VALD.\smallskip
    
    
    \item {\it  Samarium:}  HFS constants A and B on  $^{147}$\ion{Sm}{II} and $^{149}$\ion{Sm}{II} 
    are given by Masterman et al. (2003), and some more information is given by Lundquist et al. (2007).
    However, we were unable to find the configuration of upper levels in Kurucz (1993) nor in the NIST or VALD line lists. Furthermore, the 147 and 149 isotopes correspond to 14.99 and 13.82\% of the six isotopes in the solar system mixture,
    which are dominated by the even atomic numbers (Asplund et al. 2009). Therefore, we did not apply HFS for the Sm lines here, and the effect would probably be negligible.\smallskip

    \item {\it Europium} we computed the HFS for the relevant lines as they were not available in VALD.\smallskip

    \item {\it  Terbium:} the HFS splitting is given by Lawler et al. (2001c,d) and for many of the lines they are included in VALD. For four lines, not available in VALD, we computed the HFS adopting the log~gf values from Lawler et al. (2001c). Lawler et al. (2009) also gave the HFS splitting for these lines.
    \smallskip
    
    \item {\it  Dysprosium:} The isotopes $^{160,161,162,163,164}$\ion{Dy}{II} correspond to 2.33, 18.89, 25.48, 24.90 and 28.26\%, respectively, of the element constitution in the solar system mixture. Constants A and B are given by Del Papa et al. (2017). However, as with Sm, the upper level configuration is not identified for most lines. Therefore, we could not apply the HFS for the \ion{Dy}{II} lines, even though in this case the odd atomic number isotopes amount to close to 50\% of the element.\smallskip

    \item {\it  Holmium:} HFS is critical for \ion{Ho}{II}, and is available from Lawler et al. (2004).\smallskip

    \item {\it  Ytterbium:} HFS for \ion{Yb}{II} lines was adopted from Sneden et al. (2009).\smallskip

    \item {\it  Lutetium:} HFS from Den Hartog et al. (2020) is included in VALD.\smallskip
    
    \item {\it  Osmium:} HFS does not show a significant contribution to the line profile for Os, as described by Cowan et al. (2005)\smallskip

    \item {\it  Platinum:}HFS is adopted from Den Hartog et al. (2005) for the two lines.

\end{itemize}

\subsection{Hyperfine structure: Calculations}

The HFS calculations for the \ion{Ba}{II}, \ion{Eu}{II} and \ion{Tb}{II} lines were carried out using the code developed by McWilliam et al. (2013) adopting the constants A and B from Rutten (1978) and Lawler et al. (2001b), respectively.
In cases that the B constant is not provided, we adopted B\,$=$\,0.0.

The \ion{Ba}{II} line identified in the region needs to have HFS taken into account, which is not included in VALD. We adopted atomic constants from Rutten (1978). McWilliam (1998) reports HFS splitting for several \ion{Ba}{II} lines. We adopt the solar mixture isotopic fractions of 40\%, 32\%, and 28\% for $^{135}$Ba, $^{137}$Ba, and $^{138}$Ba, respectively. The HFS splitting in this case is in fact negligible. We also caution that, as extensively discussed by McWilliam (1998), the isotopic fractions could be different in old stars such as CS~31082-001, because probably no s-process contribution took place and all neutron-capture elements would have been formed by the r-process. That said, it could also be that spinstars could be responsible for forming neutron-capture elements through the s-process.


To analyse the \ion{Eu}{II} spectral lines with enough accuracy it is necessary to include the HFS, especially the transitions that include the ground configuration such as the 3688.430, 3724.930, and 3819.672~{\rm \AA} lines. We have computed the HFS shift for the \ion{Eu}{II} lines
that were not included in the VALD line lists. Eu has two stable isotopes, $^{151}$Eu and $^{153}$Eu, in the proportion of 47.9\% and 52.2\% in the solar mixture, respectively. The nuclear spin for both isotopes of \ion{Eu}{II} (151 and 153) is I\,$=$\,5/2. 
We verified that our calculations of HFS based on the code by McWilliam et al. (2013) are very similar to the HFS line splitting and respective log~gf from Ivans et al. (2006).

As noted above, most of the HFS for the selected \ion{Tb}{II} lines are available in VALD. 
We have computed the HFS for four lines that were not included in the VALD line list: \ion{Tb}{II} 3509.144, 3633.287, 3641.655 and 3899.188 {\rm \AA}. The HFS calculations for these \ion{Tb}{II} lines are somewhat simpler than for other elements given that Tb only has one isotope, $^{159}$Tb with a nuclear spin of I\,$=$\,3/2. The constants, energy levels
and log~gf values for these line transitions were adopted from Lawler et al. (2001c,d).
We verified that the line splitting from our computations, using the code by McWilliam et al. (2013) and the constants from Lawler et al. (2001c), is very similar to those given by Lawler et al. (2009).
 \smallskip
 
    
%

\section{Comments on lines}

The abundance estimates for all lines were derived from close inspection of each feature.  We now discuss each element in turn.  We note that, within the quoted uncertainties, the present results are generally compatible with previous results from Hill et al. (2002), Sneden et al. (2009), Barbuy et al. (2011), and Siqueira-Mello et al. (2013). Where small differences occur these can generally be explained by either the continuum placement and/or the adopted FWHM for convolution of the spectra in the analysis. 


\subsection{Germanium}

\begin{figure}
    \centering
    \includegraphics[width=3.3in]{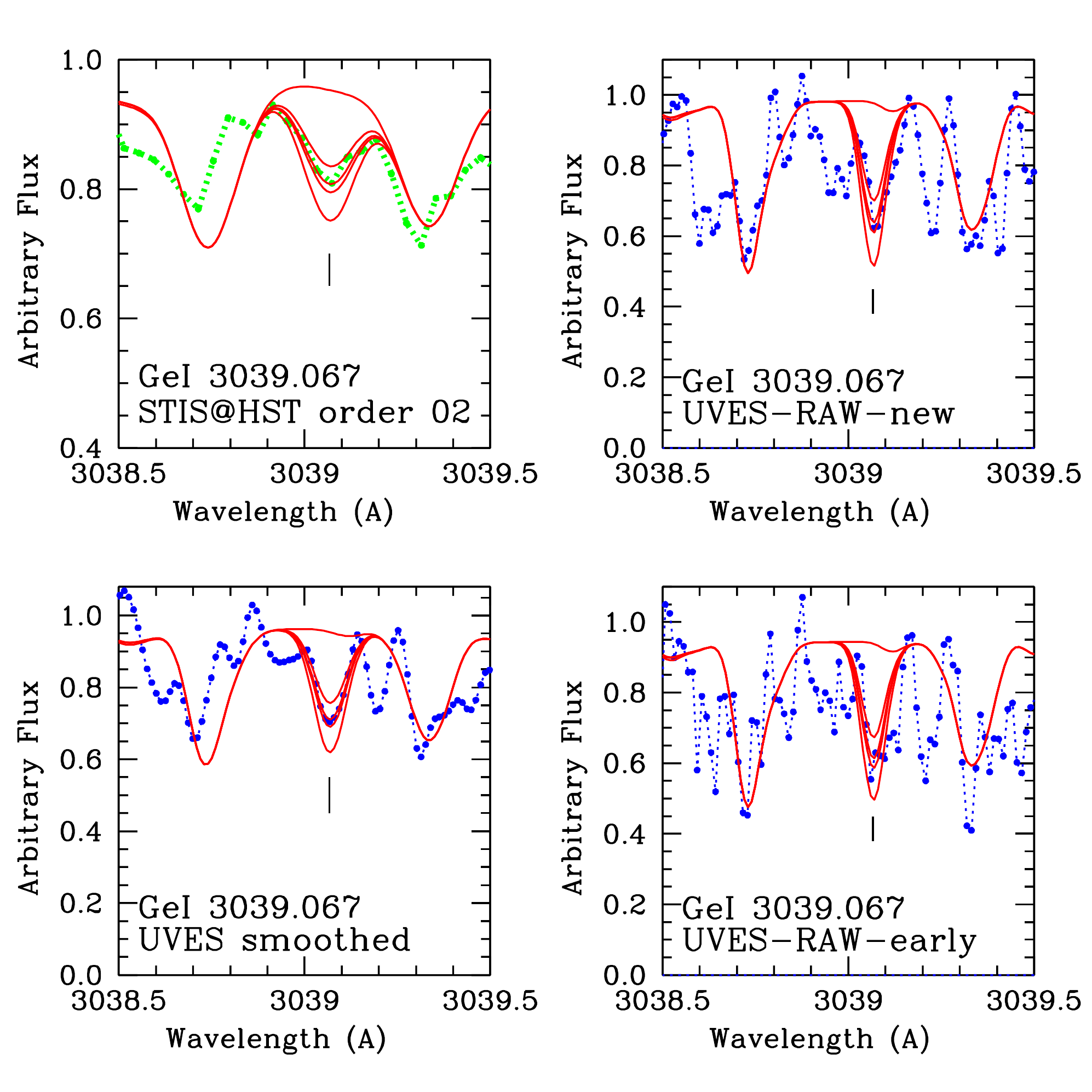}
    \caption{Fits to the \ion{Ge}{I} 3039.067 {\AA} line for the HST-STIS and UVES spectra, plotted in green and blue, respectively. The smoothed UVES data are shown in the lower-left panel, the newly-reduced data in the upper-right panel, and the previous reductions in the lower-right panel. Synthetic spectra (red lines) are shown for A(Ge)\,$=$\,none,0.71, 0.51, 0.45, and 0.31. A(Ge)\,$=$\,0.51 corresponds to [Ge/Fe]\,$=$\,0.}\label{genew}
\end{figure}

The single line of \ion{Ge}{I} 3039.067 {\rm \AA} is well-fit with A(Ge)\,$=$\,0.45 to 0.51, as shown in Fig.~\ref{genew},
therefore we adopt A(Ge)\,$=$\,0.48. The fits in the figure are for the following spectra: HST-STIS (order 2),
UVES smoothed, UVES raw (from the 2020 archival reductions), and UVES raw from the previous reductions. The lower value of A(Ge)\,$=$\,0.10 found by Siqueira-Mello et al. (2013) could be due to a different continuum placement and convolution, given that the line is weak.

\subsection{Strontium}

\begin{figure*}
    \centering
    \includegraphics[width=.9\linewidth]{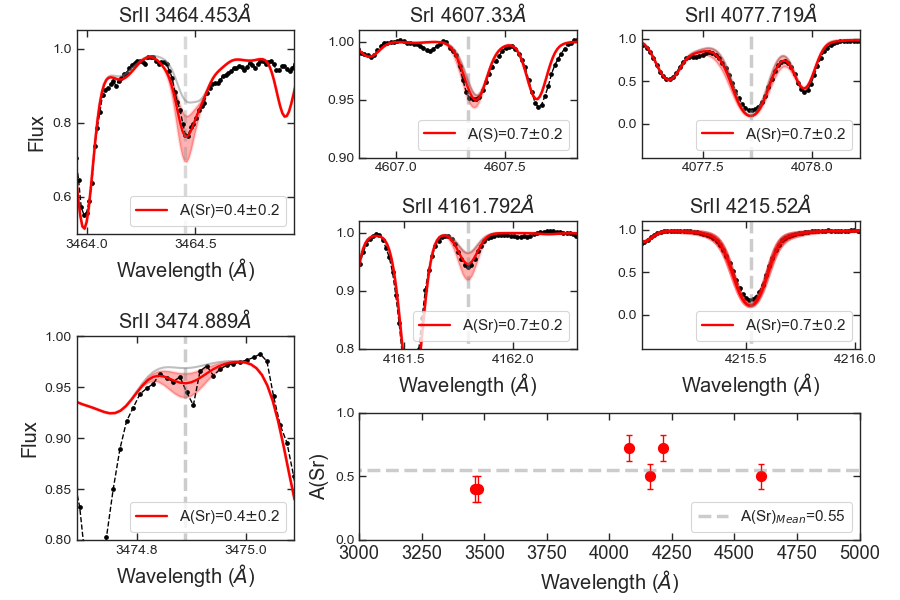}
    \caption{Fits to the \ion{Sr}{II} lines. \ion{Sr}{II} 3464.453 {\rm \AA} and 3474.889 {\rm \AA} (left-hand panels) are well fitted with A(Sr)\,$=$\,0.40 (red lines). The \ion{Sr}{II} 4077.709, 4161.792 and 4215.519 {\rm \AA} lines are well fitted with A(Sr)\,$\sim$\,0.70. The UVES spectrum is shown in black, the shaded red areas indicate an abundance variation of $\pm$\,0.2, and the grey line represents A(Sr)\,$=$\,none. The lower horizontal panel shows the mean abundance (dashed line) from the six lines, with the variation between lines within the error bars showing systematic uncertainties of $\sigma_{\rm parm}(\rm Sr)=$~0.1.}\label{Sr}
\end{figure*}

In Fig.~\ref{Sr} we show the fit of the near-UV and traditional diagnostic lines of \ion{Sr}{I} and \ion{Sr}{II}. The analysis of Sr is difficult because it has only a few useful lines in the near-UV. The \ion{Sr}{II} 3464.453 and 3474.889~{\rm \AA} lines appear too strong using the value of A(Sr)\,$=$\,0.72 from Hill et al. (2002), and are well fit with A(Sr)\,$=$\,0.40. The
\ion{Sr}{II} 3474.889~{\rm \AA} line is weak and the \ion{Sr}{II}~3464.453 {\rm \AA} line is potentially a suitable diagnostic line, except that its log~gf value is indicated in NIST as not precise, as well as having a blend with a \ion{Sm}{II} line (for which we adopted our derived value of A(Sm)\,$=$\,$-$0.54 to investigate the blend).

The lines employed by Hill et al. (2002) were more traditional for the derivation of the Sr abundance: \ion{Sr}{II} 4077.709,
4161.792, and 4215.519~{\rm \AA}. We confirmed that these lines give a Sr abundance of A(Sr)\,$=$\,0.70, including also the \ion{Sr}{I} 4607.33~{\rm \AA} line. We therefore adopted a mean value between the UV lines and the lines near 4000~{\rm \AA}, of A(Sr)\,$=$\,0.55. Note that for \ion{Sr}{II} 4077.709~{\rm \AA} we adopted the revised log~gf\,$=$\,0.148 from Roederer et al. (2022b). Fig.~\ref{Sr} shows the fits to all the Sr lines considered.


\subsection{Yttrium}

The 22 \ion{Y}{II} lines give a mean of A(Y)\,$=$\,$-$0.37\,$\pm$\,0.17. This is compatible with the result of A(Y)\,$=$\,$-$0.23 from Hill et al. (2002), who had four lines in common with our analysis, as well as six further lines in the 4300-5300~{\rm \AA} region.

\subsection{Zirconium}

The 72 \ion{Zr}{II} lines give a mean of A(Zr)\,$=$\,0.33\,$\pm$\,0.15, somewhat lower than the previous values of
A(Zr)\,$=$\,0.43 from Hill et al. (2002) and A(Zr)\,$=$\,0.55 from Siqueira-Mello et al. (2013).

Given that the higher value from Siqueira-Mello et al. (2013) was derived from the STIS spectra, we inspected the \ion{Zr}{II} lines they used, which are in STIS orders 4, 5, 6, 7, 9, 10, and 11. 
Most of the lines in the region 2600-3000 ~{\rm \AA} are blended and not very sensitive to the Zr abundance; the more reliable lines
are  \ion{Zr}{II} 2916.626~{\rm \AA} 
 and  \ion{Zr}{II} 2962.673~{\rm \AA} that give
  A(Zr)=0.43 and 0.33 respectively.
Taking into account the 72 lines, including the two best lines in the extended STIS region, we obtain a mean value of A(Zr)\,$=$\,0.33\,$\pm$\,0.15.

\subsection{Niobium}

The \ion{Nb}{II} 3028.433 {\rm \AA} line, located at the edge of the STIS order 2 spectrum, is noisy. The region
was approximately fit, but the line itself appears in the calculation as a blend with the next line shortwards of it, giving problems with fitting the continuum level as well.

For other lines, \ion{Nb}{II} 3191.093 {\rm \AA} is fitted with A(Nb)\,$=$\,$-$0.75 and \ion{Nb}{II} 3215.591 {\rm \AA} with A(Nb)\,$=$\,$-$0.65 (close to A(Nb)\,$=$\,$-$0.55 from Hill et al. 2002). \ion{Nb}{II} 3225.475 {\rm \AA} appears as a strong line, possibly due to blends and was therefore not taken into account in the mean. From these considerations, we adopt A(Nb)\,$=$\,$-$0.65.

\subsection{Molybdenum}

The unique \ion{Mo}{I} 3864.103 {\rm \AA} is well-fit with A(Mo)\,$=$\,$-$0.21, close to the previous value
of A(Mo)\,$=$\,$-$0.11\,$\pm$\,0.13 from three lines in the STIS region by Siqueira-Mello et al. (2013).
Note that this line is blended with a CN feature, but in this star the CN lines are weak.

\subsection{Ruthenium}

The \ion{Ru}{I} 3436.736, 3498.942 and 3728.025~{\rm \AA} lines are well fit with A(Ru)\,$=$\,0.15 to 0.20. The lines at 3798.898 and 3799.349~{\rm \AA} are weak and blended, but also compatible with A(Ru)\,$=$\,0.20. We also inspected the \ion{Ru}{I} 2874.988~{\rm \AA} line observed with STIS which, although very weak, is compatible with A(Ru)\,$=$\,0.25.
We therefore adopt A(Ru)\,$=$\,0.18 from the five lines in Table~\ref{Alllines}, which is  compatible with A(Ru)\,$=$\,0.36 from Hill et al. (2002) within the uncertainties (see Table~1), and lower than the result of A(Ru)\,$=$\,0.65 from Siqueira-Mello et al. (2013).

\subsection{Rhodium}

\ion{Rh}{I} 3396.819 and 3700.907~{\rm \AA} are rather weak and on the bluewards wing of stronger lines, but are well fit with A(Rh)\,$=$\,$-$0.62 and $-$0.42, respectively, whereas \ion{Rh}{I} 3434.885~{\rm \AA}  gives A(Rh)\,$=$\,$-$0.57. The \ion{Rh}{I} 3692.358~{\rm \AA} line is weak and on the redwards wing of a stronger (unidentified) line, and is compatible with A(Rh)\,$=$\,$-$0.42. A mean value of A(Rh)\,$=$\,$-$0.51\,$\pm$\,0.09 is close to the abundance of A(Rh)\,$=$\,$-$0.42 from Hill et al. (2002). 

\subsection{Palladium}

The five lines give a mean of A(Pd)\,$=$\,$-$0.21\,$\pm$\,0.07, somewhat lower than the value of A(Pd)\,$=$\,$-$0.05 from Hill et al. (2002). There is good line-by-line agreement between the estimates from the raw (non-smoothed) UVES and Keck spectra.

\subsection{Silver}

The two lines give a mean of A(Ag)\,$=$\,$-$0.94\,$\pm$\,0.07, in very good agreement with A(Ag)\,$=$\,$-$0.95 from Hill et al. (2002), and compatible with the value A(Ag)\,$=$\,$-$1.03 from Siqueira-Mello et al. (2013). The results from the smoothed and raw UVES data and the Keck spectra are all very similar.

\subsection{Cadmium}
The sole Cd line is hardly detectable, and moreover, is immersed within OH lines. Only a very small change in the feature takes place by changing its abundance by 0.2~dex, and therefore the Cd abundance cannot be derived.

\subsection{Tin}

\begin{figure}
    \centering
    \includegraphics[width=3.3in]{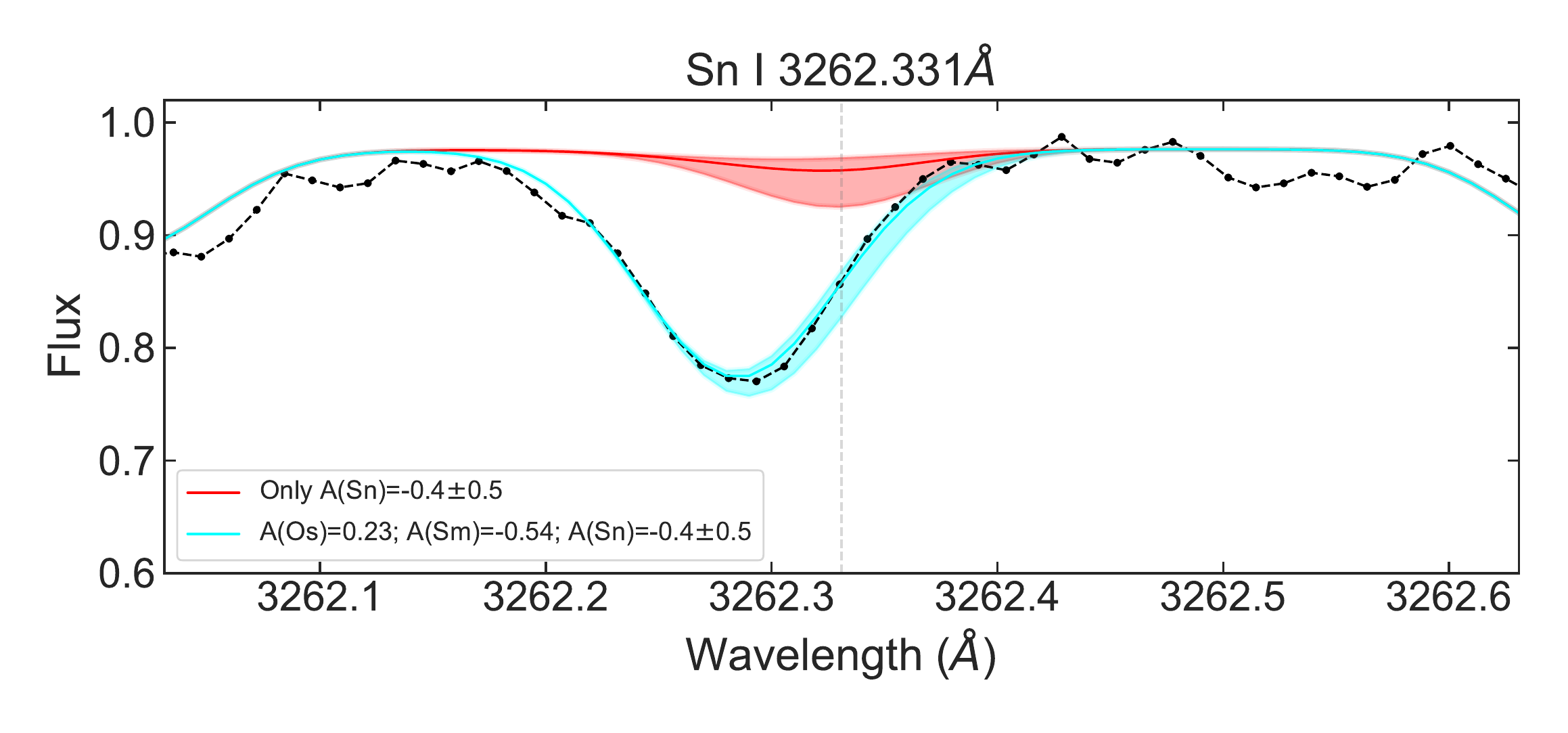}
    \caption{Fit to the \ion{Sn}{I} 3262.331 {\rm \AA} line that includes calculations with A(Sn)\,$=$\,$-$0.90,$-$0.40,$-$0.20, 0.0. The red region shows the models including only Sn, while the cyan region includes blending from Os and Sm, with A(Os)\,$=$\,0.0 and A(Sm)\,$=$\,$-$0.51, respectively.}\label{SN}
\end{figure}

The one \ion{Sn}{I} line, at 3262.331~{\rm \AA}, is blended with \ion{Sm}{II} 3262.27~{\rm \AA} and \ion{Os}{I} 3262.29~{\rm \AA}, as pointed out by Sneden et al. (2003) and shown in Fig.~\ref{SN}. The blend is well fit with A(Os)\,$=$\,0.0 (where the Os line is the stronger feature), A(Sm)\,$=$\,$-$0.51, and A(Sn)\,$=$\,$-$0.40 or [Sn/Fe]\,$=$\,$+$0.5. We add though that the Sn abundance in \ion{Sn}{I} 3262.331~{\rm \AA} has a small impact in the overall feature, so we only adopted an upper limit on the Sn abundance, of A(Sn)\,$<$\,$-$0.40. Sn is mostly produced by the s-process in the solar mixture, and in CS~31082-001 it is probably produced in a small contribution from the r-process, which in principle explains that this element is not much enhanced in this star. There are two other Sn lines in this region, but they are undetectable -- the \ion{Sn}{I} 3655.790~{\rm \AA} line has its feature strongly overlapping with a \ion{Ce}{II} line, while the 3801.011~{\rm \AA} line is too weak and so presents no detectable feature for A(Sn)\,$=$\,$-$0.40.

\subsection{Barium}

\begin{figure*}
    \centering
    \includegraphics[width=.9\linewidth]{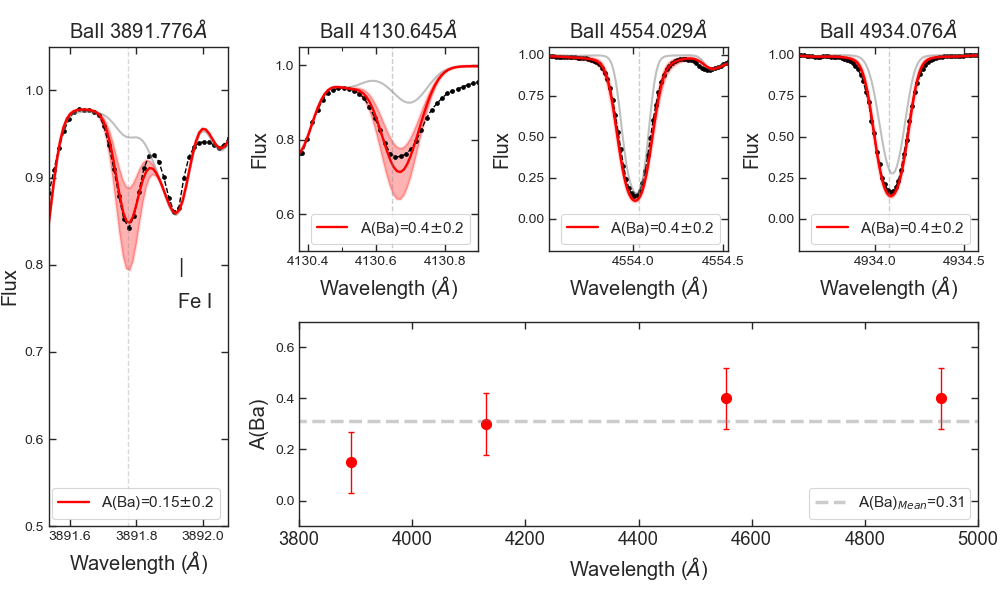}
    \caption{Fits to the \ion{Ba}{II} 3891.776 {\rm \AA} line in the near-UV and to the more traditional optical \ion{Ba}{II} 4130.645, 4554.229 and 4934.076 {\rm \AA} lines. The UVES spectrum is shown in black, the shaded red areas indicate an abundance variation of $\pm$\,0.15, and the grey line represents A(Ba)\,$=$\,none. The lower horizontal panel shows the mean abundance (dashed line) from the four lines, with the variation between lines with the error bars showing systematic uncertainties of $\sigma_{\rm parm}$(\rm Ba)\,$=$\,0.12.}\label{BA}
\end{figure*}

The sole useful \ion{Ba}{II} line is at 3891.776~{\rm \AA} and is blended with a \ion{Fe}{I} line. This is well fit with A(Ba)\,$=$\,0.15, as shown in Fig.~\ref{BA}, where we also show fits to more traditional Ba lines in the range 4000-5000~{\rm \AA}, i.e. \ion{Ba}{II} 4130.645, 4554.229, and 4934.076~{\rm \AA}. A mean value of A(Ba)\,$=$\,0.40 is obtained, in agreement with the mean value from Hill et al. (2002).

\subsection{Lanthanum}

The mean of the six lines gives A(La)\,$=$\,$-$0.73\,$\pm$\,0.07, only somewhat lower than A(La)\,$=$\,$-$0.60 from Hill et al. (2002). The results from fits with the three sets of spectra are all very similar.

\subsection{Cerium}

The 19 lines give a mean of A(Ce)\,$=$\,$-$0.41\,$\pm$\,0.11, only slightly lower than A(Ce)\,$=$\,$-$0.31 from Hill et al. (2002). The exceptions are the \ion{Ce}{II} 3263.885 and 3984.671~{\rm \AA} lines that are not well fit due to blends and these lines were therefore not considered. The \ion{Ce}{II} 3999.237~{\rm \AA} line was only used in the case of the raw (non-smoothed) UVES spectrum.

\subsection{Praseodymium}

The triplet lines are well fit by a mean of A(Pr)\,$=$\,$-$0.94\,$\pm$\,0.02, close to the value of A(Pr)\,$=$\,$-$0.86 from Hill et al. (2002).

\subsection{Neodymium}

A mean of A(Nd)\,$=$\,$-$0.33\,$\pm$\,0.09 is found from 23 lines, somewhat lower than the abundance of A(Nd)\,$=$\,$-$0.13 from Hill et al. (2002). The exceptions are: \ion{Nd}{II} 3285.085 {\rm \AA} which is faint and just longwards of a stronger line (with a fit to the blended feature giving a lower Nd abundance of A(Nd)\,$=$\,$-$0.53), and \ion{Nd}{II} 3334.465~{\rm \AA} which is also blended; these two lines are not included in the mean. The abundance of A(Nd)\,$=$\,$-$0.21 from Siqueira-Mello et al. (2013) is compatible within the uncertainties.

\subsection{Samarium}

The mean of A(Sm)\,$=$\,$-$0.54 is very close to the value of A(Sm)\,$=$\,$-$0.51 from Hill et al. (2002).

\subsection{Europium}

As explained in Sect. 3.2, the Eu lines require HFS, which we included in our calculations. We obtained a mean of A(Eu)\,$=$\,$-$0.93\,$\pm$\,0.07, somewhat lower than A(Eu)\,$=$\,$-$0.76 from Hill et al. (2002).

\subsection{Gadolinium}

The 39 measurable Gd lines give a mean of A(Gd)\,$=$\,$-$0.45\,$\pm$\,0.10, somewhat lower than A(Gd)\,$=$\,$-$0.27 from Hill et al. (2002). For the shortest wavelength lines (\ion{Gd}{II} 3032.844 and 3034.051~{\rm \AA}), 
the STIS spectrum is well-fit with A(Gd)\,$=$\,$-$0.47, and would need a higher Gd abundance by about 0.15 dex to fit the UVES spectrum. The \ion{Gd}{II} 3360.712~{\rm \AA} line is blended with NH lines, and \ion{Gd}{II} 3482.607~{\rm \AA} is blended with both \ion{Co}{I} 3482.634~{\rm \AA} and molecular features, and could not be used.

\subsection{Terbium}

Most of the 12 Tb lines are well fitted with A(Tb)\,$=$\,$-$1.26 as derived by Hill et al. (2002), and this applies to the three sets of spectra. The \ion{Tb}{II} 3874.168~{\rm \AA} line appears as an asymmetry on the redwards wing of a stronger line. We also inspected the \ion{Tb}{II} 2934.802~{\rm \AA} line observed with STIS; this is a weak line, requiring a lower Tb abundance from the order 5 spectrum and a slightly higher value from order 4. Therefore, the higher abundance of A(Tb)\,$=$\,$-$0.50 derived by Siqueira-Mello et al. (2013) is not justified, and we find a mean value of A(Tb)\,$=$\,$-$1.22\,$\pm$\,0.10.

\subsection{Dysprosium}

Except for the shorter wavelength and weak \ion{Dy}{II} 3026.160~{\rm \AA} line measured in the STIS spectrum, 
the other 27 Dy lines give a mean of A(Dy)\,$=$\,$-$0.25\,$\pm$\,0.08, very close to the value of A(Dy)\,$=$\,$-$0.21 reported by Hill et al. (2002). Therefore, the somewhat higher value of A(Dy)\,$=$\,$-$0.12 from Siqueira-Mello et al. (2013) is compatible within the uncertainties.

\subsection{Holmium}

\begin{figure*}
    \centering
    \includegraphics[width=5.5in]{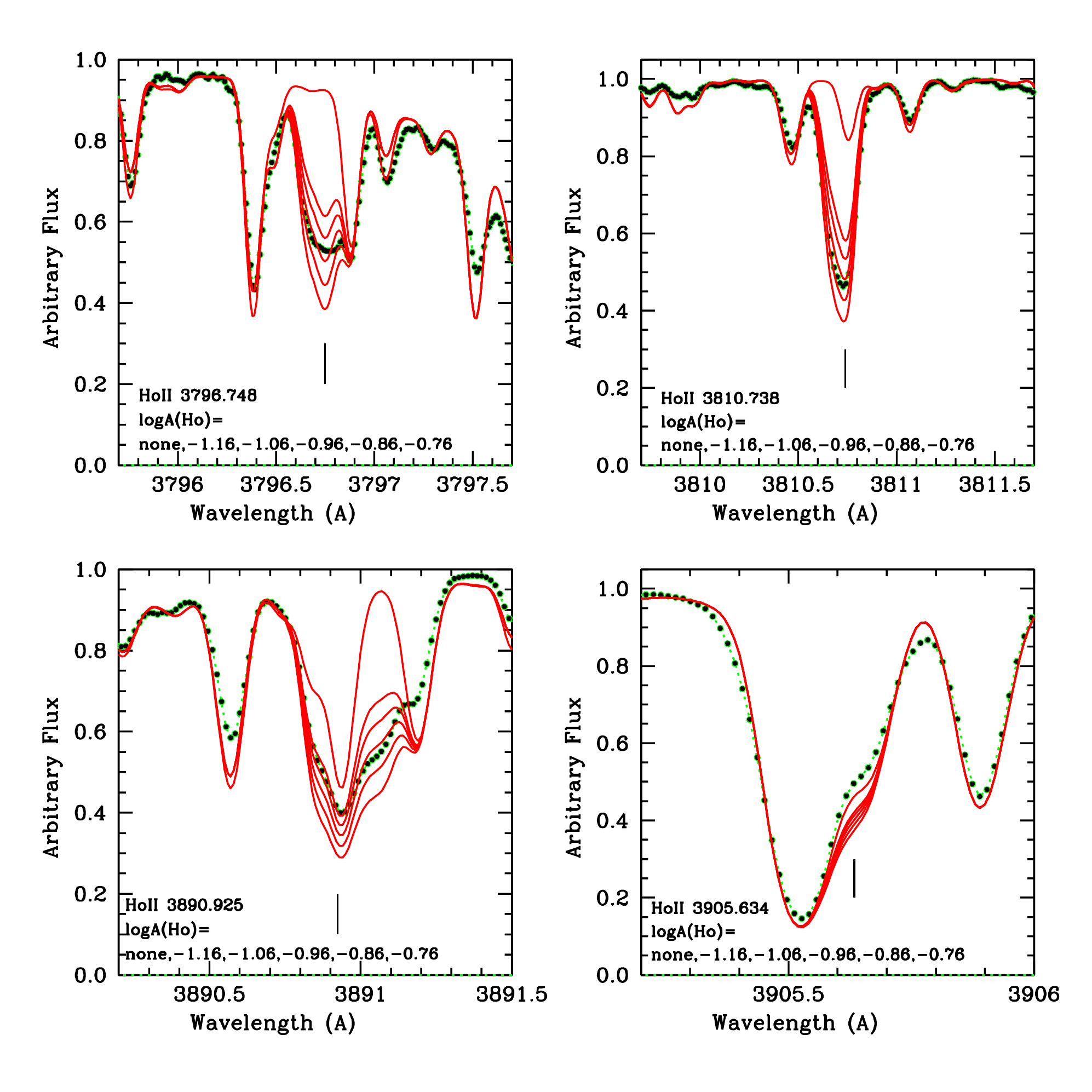}
    \caption{Fits to the \ion{Ho}{II} 3796.748, 3810.738, 3890.925 and 3905.634 {\AA} lines in the UVES spectrum (black points), in which the synthetic spectra (red lines) were computed for A(Ho)\,$=$\,none, $-$1.16, $-$1.06, $-$0.96, $-$0.86, and $-$0.76.}\label{ho}
\end{figure*}

We obtain a mean of A(Ho)\,$=$\,$-$0.98\,$\pm$\,0.06. The fits to the four lines are shown in Fig.~\ref{ho}, but the \ion{Ho}{II} 3890.925 and 3905.634~{\rm \AA} lines shown in the two lower panels are heavily blended and are not considered in calculation of the final value. 

\subsection{Erbium}

We find a mean of A(Er)\,$=$\,$-$0.38\,$\pm$\,0.10, somewhat lower than A(Er)\,$=$\,$-$0.27 from Hill et al. (2002).

\subsection{Thulium}

Most of the 10 lines available are well fit with A(Tm)\,$=$\,$-$1.24 as derived by Hill et al. (2002). We find a mean of A(Tm)\,$=$\,$-$1.32\,$\pm$\,0.16. The \ion{Tm}{II} 3362.615~{\rm \AA} line is blended with NH lines and therefore not reliable
as a diagnostic of the Tm abundance.

\subsection{Ytterbium}

\begin{figure}
    \centering
    \includegraphics[width=3.3in]{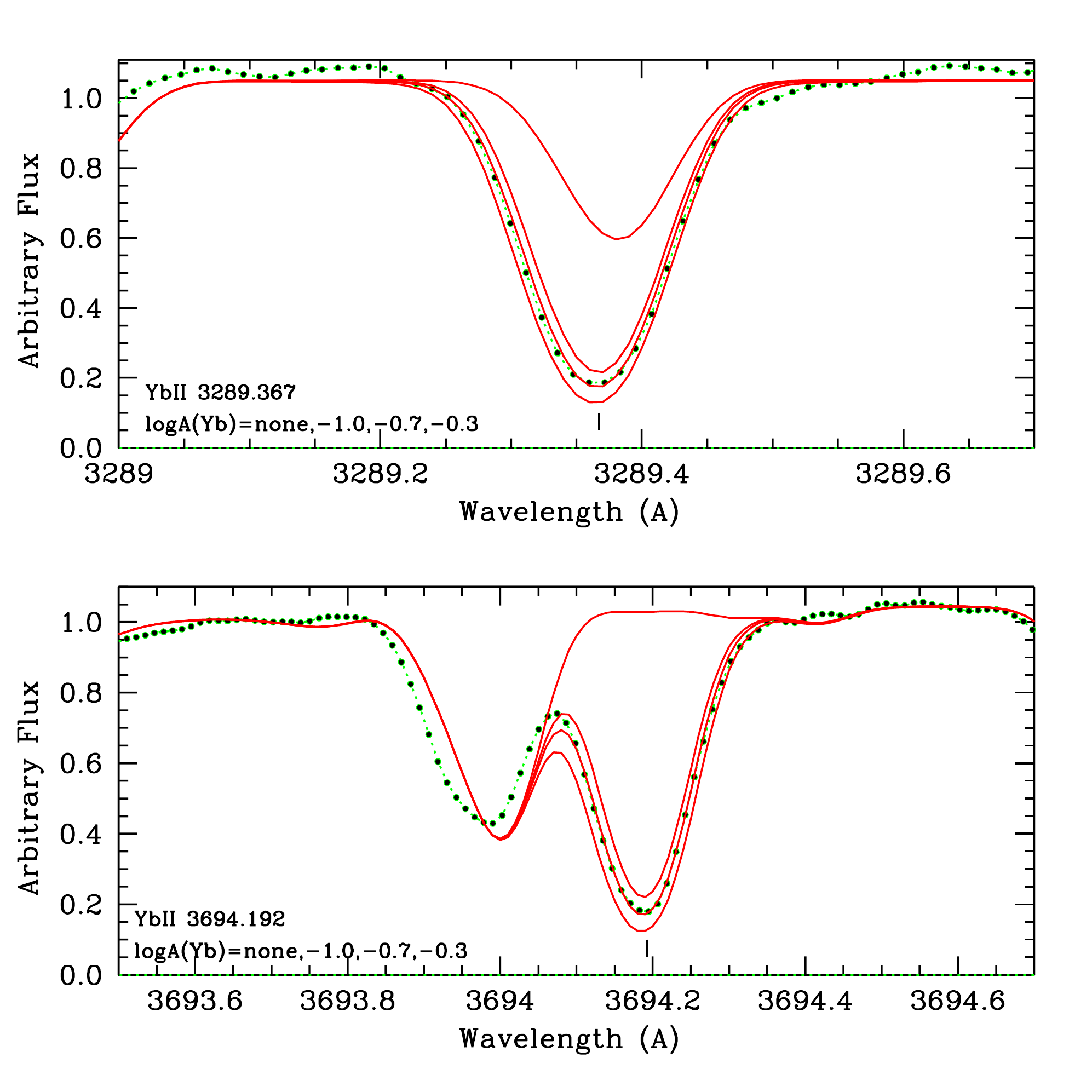}
    \caption{Fits to the \ion{Yb}{II} lines in the UVES spectrum (black points), in which the synthetic spectra (red lines) are computed for A(Yb)\,$=$\,none, $-$1.0, $-$0.7, $-$0.3.}\label{YB}
\end{figure}

The two lines of Yb are well-fit with A(Yb)\,$=$\,$-$0.70, somewhat lower than the value of A(Yb)\,$=$\,$-$0.41 from Sneden et al. (2009); the fits to the lines are shown in Fig.~\ref{YB}. The \ion{Yb}{II}~3289.367~{\rm \AA}  line is blended with \ion{V}{II} 3289.390~{\rm \AA} ($\chi_{\rm ex}$\,$=$\,1.096 eV, log~gf\,$=$\,$-$0.931) and \ion{Fe}{II} 3289.354~{\rm \AA}  ($\chi_{\rm ex}$\,$=$\,3.814 eV, log~gf\,$=$\,$-$1.620). The abundances of Fe and V are well constrained, and therefore these blends do not cause significant uncertainties.

\subsection{Lutetium}

\ion{Lu}{II} 3077.605~{\rm \AA} is effectively the only clear line for lutetium. It is well fitted with A(Lu)\,$=$\,$-$1.14, or [Lu/Fe]\,$=$\,$+$1.0. The \ion{Lu}{II} 3397.066~{\rm \AA} line is on the redwards wing of another line, and the asymmetry caused by the Lu line is well fit with A(Lu)\,$=$\,$-$1.14 as well. \ion{Lu}{II} 3472.477~{\rm \AA} is on the bluewards wing of a strong line, and is very weak, and not useful. Finally, \ion{Lu}{II} 3554.516~{\rm \AA} line is weak and blended, and shows no sensitivity to the Lu abundance and we do not use it further here.

\subsection{Hafnium}    

From the seven lines of Hf we obtain a mean of A(Hf)\,$=$\,$-$0.88\,$\pm$\,0.05, somewhat lower than A(Hf)\,$=$\,$-$0.59 derived by Hill et al. (2002). The \ion{Hf}{II} 3793.379~{\rm \AA} line is less reliable than the others because it is in the wing of another stronger line, and many of the nearby lines are not well-reproduced. 

\subsection{Osmium}

We have fitted the \ion{Os}{I} 3018.036 and 3058.655~{\rm \AA} lines from the STIS order 3 and 2 spectra, respectively. The \ion{Os}{I} 3018.036~{\rm \AA} is faint and blended, but can be fitted with A(Os)\,$=$\,0.10, compatible with A(Os)\,$=$\,$-$0.07 from Barbuy et al. (2011).  The \ion{Os}{I} 3058.655~{\rm \AA} line in order 3 of the STIS spectrum appears to be far more reliable, with a higher S/N. We obtain a mean of A(Os)\,$=$\,0.23\,$\pm$\,0.16.

\subsection{Iridium}

Adopting the abundance A(Ir)\,$=$\,$+$0.20 from Hill et al. (2002), the \ion{Ir}{I} 3047.158~{\rm \AA} line is weak but well fit, as well as \ion{Ir}{I} 3513.648~{\rm \AA} which is located within a set of other lines. We find a mean of A(Ir)\,$=$\,0.10\,$\pm$\,0.12, compatible with A(Ir)\,$=$\,0.20 from Barbuy et al. (2011) and Hill et al. (2002).

 
\subsection{Platinum}

\begin{figure*}
    \centering
    \includegraphics[width=5.5in]{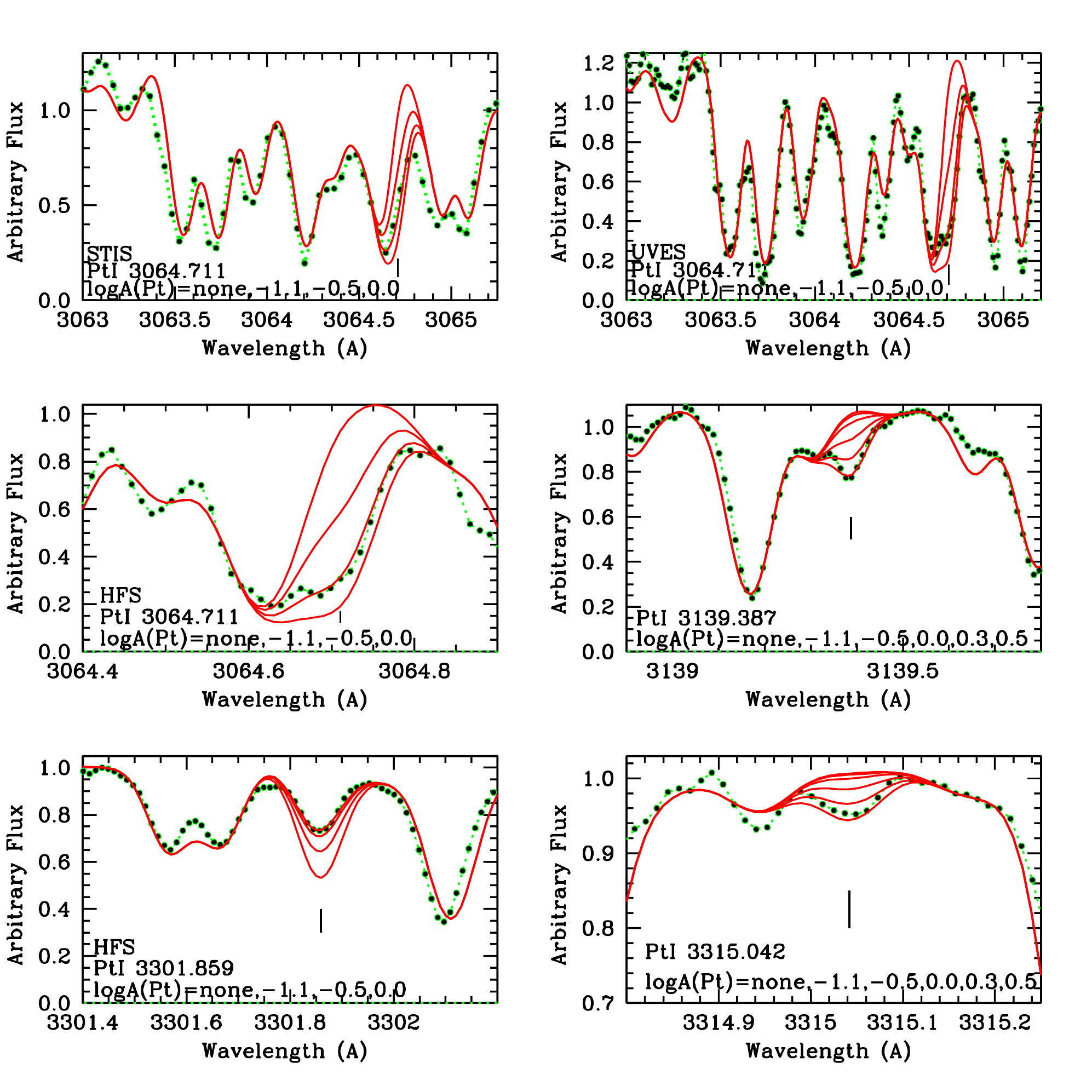}
    \caption{Fits to the \ion{Pt}{I} lines. The upper left- and right-hand panels show fits to the 3064.711 {\AA} line in the STIS and UVES data, respectively; the middle-left panel shows a zoom-in on the region for the UVES data.  The remaining panels show the fits to the other three lines.  The synthetic spectra shown (red lines) are computed for A(Pt)\,$=$\,none, $-$1.1, $-$0.5, 0.0, $+$0.3, $+$0.5.}\label{ptuves}
\end{figure*}

Our fits to the Pt lines are shown in Fig.~\ref{ptuves}. For the \ion{Pt}{I} 3064.711 and 3301.859~{\rm \AA} lines we adopted
the HFS splitting from Den Hartog et al. (2005). The \ion{Pt}{I} 3064.711~{\rm \AA} line, observed both with UVES and STIS (order 2), is blended with another line, but is clearly seen and gives A(Pt)\,$=$\,$-$0.50. This is the most reliable indicator, given that it has HFS included, and is not prohibitively blended. The \ion{Pt}{I} 3301.859~{\rm \AA} line gives A(Pt)\,$=$\,$-$1.1, but there is a strong blend and it is not considered further. The \ion{Pt}{I} 3139.385 and 3315.042~{\rm \AA} lines have no HFS applied, and give a much higher value of A(Pt)\,$=$\,0.50. Given the various challenges of the lines we adopt a mean value of A(Pt)\,$=$\,0.00.

\subsection{Gold}

The only \ion{Au}{I} line available in our spectra (2675.937~{\rm \AA}) is located in STIS order 11. Compared to the study from Barbuy et al. (2011), blending lines on the redward side of the \ion{Au}{I} line have now been identified as \ion{Fe}{I} 2676.015, 2676.079 and 2676.160~{\rm \AA}, with excitation potentials of 2.728, 2.949, 2.609~eV, respectively. The oscillator strengths of these lines had to be changed, from $-$2.404 to $-$1.35, $-$0.382 to $-$1.15 and $-$1.828 to $-$1.778, respectively, to fit the feature; the \ion{Fe}{I} 2676.015~{\rm \AA} line is particularly important in the blend. A revised value for the Au abundance is now A(Au)\,$=$\,$-$1.39, or [Au/Fe]\,$=$\,$+$0.6 adopting A(Au)$_{\odot}$\,$=$\,0.91 from Asplund et al. (2021). However, there remains uncertainty due to other blends superimposed on the \ion{Au}{I} line, namely, a rather weak line of \ion{Fe}{II} 2675.901~{\rm \AA} ($\chi_{\rm ex}$\,$=$\,5.823~eV, log~gf\,$=$\,$-$2.088), and lines of \ion{Co}{I} 2675.981~{\rm \AA}, ($\chi_{\rm ex}$\,$=$\,0.629~eV, log~gf\,$=$\,$-$1.625), \ion{Ta}{II} 2675.900~{\rm \AA} ($\chi_{\rm ex}$\,$=$\,0.548~eV, log~gf\,$=$\,0.870) and \ion{Nb}{II} 2675.939~{\rm \AA} ($\chi_{\rm ex}$\,$=$\,0.054~eV, log~gf\,$=$\,$-$0.730). The abundances of Co, Nb and Ta are from Cayrel et al. (2004) and Siqueira-Mello et al. (2013), although the lines of Nb and Ta are weak. Other contaminants reported by Barbuy et al. (2011) are no longer present in the revised line lists from VALD.

\subsection{Lead}

The two available lines in the 3000-4000 {\rm \AA} region are very faint and cannot be used: \ion{Pb}{I} 3639.568~{\AA} is not detectable, and \ion{Pb}{I} 3683.462~{\AA} is detectable but weak and in the wing of a very strong line, and therefore not measurable. The \ion{Pb}{I} 2833.053~{\AA} line in the STIS spectrum (with a nearby \ion{Ir}{II} 2833.241~{\AA} line) has an excitation potential of 1.453~eV and the oscillator strength from VALD of log~gf\,$=$\,1.924 had to be corrected to $-$0.30 to fit the line. This UV \ion{Pb}{I} line is well fit with A(Pb)\,$=$\,$-$0.65, in agreement with Barbuy et al. (2011), and only 0.1~dex lower than the value of A(Pb)\,$=$\,$-$0.55 given by Plez et al. (2004) from \ion{Pb}{I} 4057.8~{\rm \AA}.

\subsection{Bismuth}

\begin{figure*}
    \centering
    \includegraphics[width=5.5in]{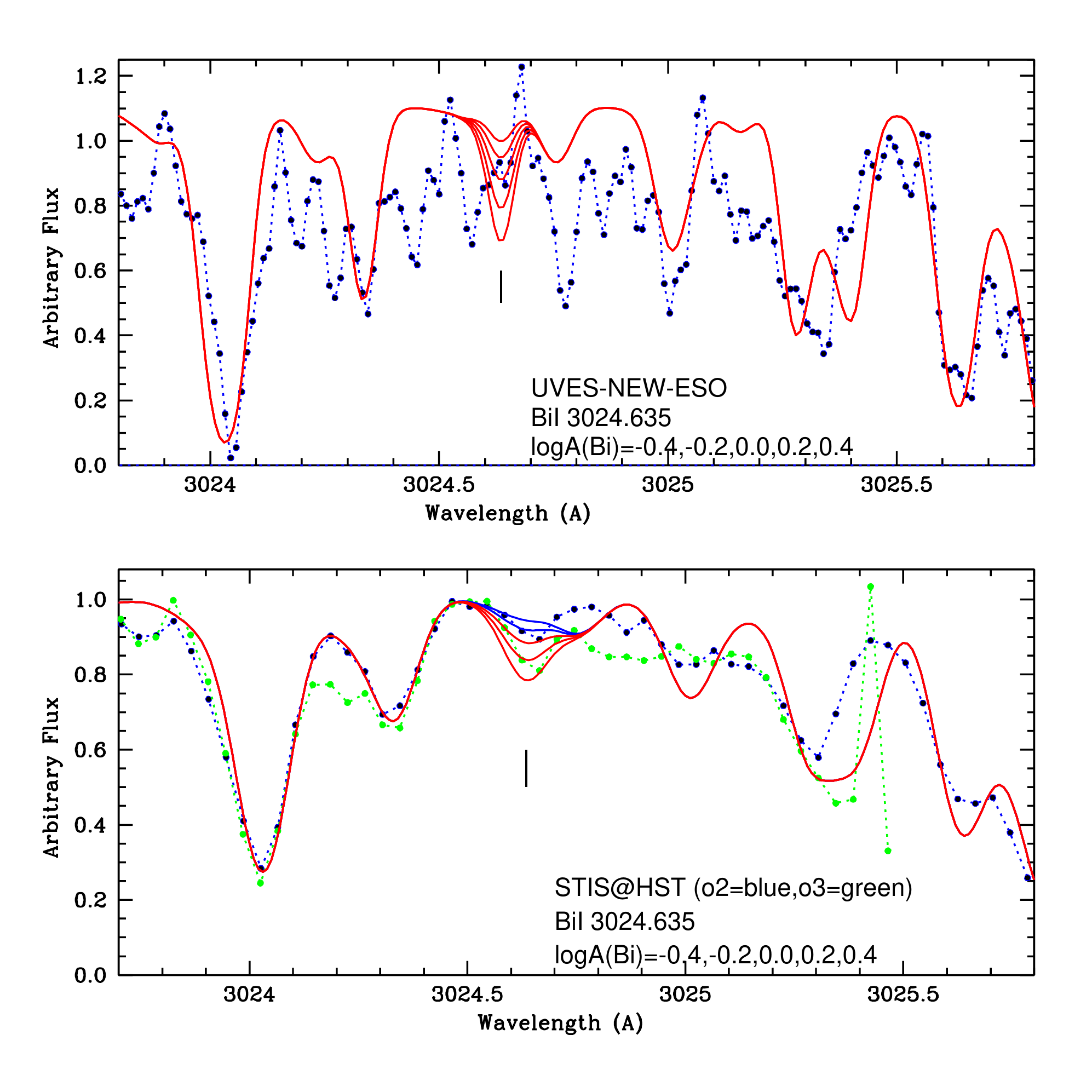}
    \caption{Fits to the \ion{Bi}{I} 3024.64 {\AA} line for the UVES and STIS spectra.  The UVES spectrum is shown by the blue dotted line in the upper panel, and the STIS order 2 and 3 spectra are shown by the blue and green dotted lines, respectively. The synthetic spectra shown
    (red lines) are computed for A(Bi)\,$=$\,none, $-$0.4, $-$0.2, 0.0, $+$0.2, $+$0.4.}\label{bisuvesnew}
\end{figure*}

The \ion{Bi}{I} 3024.635~{\rm \AA} line was observed both with UVES and with STIS (in orders 2 and 3), and the fits are shown in Fig.~\ref{bisuvesnew}. In our fits we tried to adopt log~gf\,$=$\,$-$0.15 as reported in NIST and measured by Wahlgren et al. (2001). This value contrasts with the log~gf\,$=$\,1.35 adopted by Barbuy et al. (2011), from measurements by Andersen et al. (1972) and which had been adopted by VALD and included in the Kurucz line lists. However, the smaller value leads to A(Bi)\,$=$\,1.5 and [Bi/Fe]\,$=$\,$+$3.75, that is far too over-enhanced and probably not realistic, and we therefore retained the previous log~gf\,$=$\,1.35 in our calculations. Our fit to the order~2 STIS spectrum (blue line in the figure) gave A(Bi)\,$=$\,$-$0.2, and is more reliable than the fit to order 3, which give A(Bi)\,$=$\,$+$0.2 (green line). In the higher-resolution UVES spectrum (upper panel), despite the noise, it appears to be compatible with A(Bi)\,$=$\,$-$0.2. As pointed out by Barbuy et al. (2011), the Bi abundance is an important calibration point for zero-age r-process abundance distribution models (Schatz et al. 2002).

\subsection{Thorium}

Six lines are weak but measurable and give a mean of A(Th)\,$=$\,$-$1.04\,$\pm$\,0.08, very close to the value of A(Th)\,$=$\,$-$0.98 from Hill et al. (2002). The \ion{Th}{II} 3180.194~{\rm \AA} line is not only weak but also blended with a strong line, and the overall feature is very insensitive to the Th abundance. We also note that the \ion{Th}{II} 3675.567~{\rm \AA} line is well fit, but that the surrounding lines are not well reproduced with the present line list, even if this does not affect the Th line.

\subsection{Uranium}

Cayrel et al. (2001) derived A(U)\,$=$\,$-$1.6, and Hill et al. (2002) revised the value to A(U)\,$=$\,$-$1.92, as a result of a new measurement of the line oscillator strength from the UVES spectra. Barbuy et al. (2011) adopted this latter value, which we also adopt here.

\subsection{Uncertainties}

The final abundances in Table \ref{heavy} correspond to the estimates from the UVES raw (non-smoothed) spectra. We consider this as the best quality spectrum, taking into account the excellent sampling of these data, with a pixel scale of 0.0147 {\rm \AA}/pix and $\geq$5 pixels per resolution element. Table \ref{heavy_error} gives the uncertainties from a) the fits to lines in the STIS spectra for wavelengths $<$3070 {\rm \AA}, and otherwise the smoothed UVES spectrum centered at 3400 {\rm \AA}; b) fits to the Keck spectra; c) fits to the raw (non-smoothed) UVES spectrum. 

The statistical uncertainty was computed through the standard deviation formula:

$$  \sigma= \sum_{i=1}^{N} \sqrt{\frac{(X_i-\mu)^2}{N}},    $$

and added to this are the systematic uncertainties resulting from stellar parameters (column 3 of Table~\ref{heavy_error}). 
For most of the elements we adopted the errors in stellar parameters as given in Table~6 of Hill et al. (2002), and also took into account the errors computed by Siqueira-Mello et al. (2013) and Sneden et al. (2009). For the remaining elements that had not been analysed before we estimated the error by computing models with $\Delta$T\,$=$\,$+$100~K, $\Delta$log~g\,$=$\,$+$0.3~dex, $\Delta$v$_{\rm t}$\,$=$\,$+$0.2~km\,s$^{-1}$. Table~\ref{errors} reports the uncertainties for elements not studied previously.

The mean final abundances in Table~\ref{heavy} were obtained from the mean value among the spectral lines considered in Table \ref{Alllines}, in some cases excluding the outliers. For some elements the previous abundance estimates from the STIS spectra were kept as the final value as the quality of the observed lines is better for some lines in these data.

\section{Discussion}

\setlength{\tabcolsep}{10pt}
\begin{table*}
\centering 
\scalefont{0.88}
\caption{Summary of heavy-element abundances for CS~31082-001 derived in this study, A(X)$_{\rm present}$ and [X/Fe]$_{\rm present}$, compared to those from previous works as follows: 1) Hill et al. (2002), 2) Plez et al. (2004), 3) Barbuy et al. (2011), 4) Siqueira-Mello et al. (2013), 5) Sneden et al. (2009). Solar abundances in the third column are adopted from Asplund et al. (2021).
The uncertainties ($\sigma_{\rm [X/Fe]}$) are computed considering the statistical ($\sigma_{\rm fit}$) and systematic 
uncertainties that come from the fit of the number of lines (N) for each element, where the systematic 
uncertainties come from the stellar parameters (col.~3 in Table~2) and the derivations from the UVES spectra 
(col.~10 in Table~2). Notation "1st *s" means "First Stars".}\label{heavy}
\begin{tabular}{lccccccccc} 
\hline\hline             
Element  &  Z  &  A(X)$_{\odot}$$_{\rm adopted}$  & A(X)$_{\rm VLT,Keck}$ $^{\rm 1st *s}$ 
& A(X)$_{\rm HST}$ $^{\rm 1st *s}$ & N & A(X)$_{\rm present}$  &  [X/Fe]$_{\rm present}$ & $\sigma_{\sqrt{fit^2+param^2}}$ \\
\hline  
\hline     
Ge & 32 &3.62     & \ldots                                      & \,$+$0.10$^{4}$     & 1     & \pp0.48     &$-$0.24\pp      & 0.16   \\  
Sr & 38 &2.83     & +0.72$^{1}$                                 & \ldots         & 2     & \pp0.55  & 0.62  &         0.16  \\  
Y  & 39 &2.21     & $-$0.23$^{1}$, $-$0.15$^{4}$                & \ldots         & 22    &$-$0.37  & 0.32        & 0.21  \\  
Zr & 40 &2.59     & +0.43$^{1}$                                 & \,+0.55$^{4}$    & 66    & \pp0.26     & 0.57        & 0.19  \\  
Nb & 41 &1.47     & $-$0.55$^{1}$                               & $-$0.52$^{4}$  & 3     &$-$0.65  & 0.78        & 0.14  \\  
Mo & 42 &1.88     & \ldots                                      & $-$0.11$^{4}$  & 1     &$-$0.21  & 0.81        & 0.09   \\  
Ru & 44 &1.75     &  +0.36$^{1}$                                & \,+0.65$^{4}$    & 5     & \pp0.18     & 1.33        & 0.16  \\  
Rh & 45 &0.78     & $-$0.42$^{1,4}$                             & \ldots         & 3     &$-$0.51  & 1.61        & 0.19  \\  
Pd & 46 &1.57     & $-$0.05$^{1}$, $-$0.09$^{4}$                & \ldots         & 5     &$-$0.21  & 1.12        & 0.18  \\  
Ag & 47 &0.96     & $-$0.81$^{1}$, $-$0.84$^{4}$                &  \ldots        & 2     &$-$0.94  & 1.00        & 0.18  \\  
Sn & 50 &2.02     & \ldots                                      &\ldots          & 1     &$<$$-$0.40  & $<$0.48\pp        & 0.22   \\  
Ba & 56 &2.27     & +0.40$^{1}$                                 & \ldots         & 1     & \pp0.40     & 1.03        & 0.15   \\  
La & 57 &1.11     & $-$0.60$^{1}$, $-$0.60$^{5}$                & \ldots         & 6     &$-$0.73  & 1.06        & 0.15  \\  
Ce & 58 &1.58     & $-$0.31$^{1,4}$,$-$0.29$^{5}$               &\ldots          & 18    &$-$0.41  & 0.91        & 0.16  \\  
Pr & 59 &0.75     & $-$0.86$^{1}$,$-$0.79$^{5}$                 &\ldots          & 3     &$-$0.94  & 1.21        & 0.12  \\  
Nd & 60 &1.42     & $-$0.13$^{1}$, $-$0.21$^{4}$,$-$0.15$^{5}$  &\ldots          & 24    &$-$0.33  & 1.15        & 0.15  \\  
Sm & 62 &0.95     & $-$0.51$^{1}$, $-$0.42$^{4,5}$              &\ldots          & 26    &$-$0.54  & 1.41        & 0.22  \\  
Eu & 63 &0.52     & $-$0.76$^{1}$,$-$0.72$^{5}$                 & $-$0.75$^{4}$  & 6     &$-$0.93  & 1.45        & 0.14  \\  
Gd & 64 &1.08     & $-$0.27$^{1}$,$-$0.21$^{5}$                 & $-$0.22$^{4}$  & 39    &$-$0.45  & 1.37        & 0.16  \\  
Tb & 65 &0.31     & $-$1.26$^{1}$,$-$1.01$^{5}$                 & $-$0.50$^{4}$  & 11    &$-$1.22  & 1.37        & 0.15  \\  
Dy & 66 &1.10     & $-$0.21$^{1}$, $-$0.12$^{4}$,$-$0.07$^{5}$  & \ldots         & 28    &$-$0.25  & 1.55        & 0.14  \\  
Ho & 67 &0.48     & $-$0.80$^{5}$                               &\ldots          & 3     &$-$0.98  & 1.44        & 0.15  \\  
Er & 68 &0.93     & $-$0.27$^{1}$,$-$0.30$^{5}$                 & $-$0.20$^{4}$  & 21    &$-$0.38  & 1.59        & 0.17  \\  
Tm & 69 &0.11     & $-$1.24$^{1}$, $-$1.18$^{4}$,$-$1.15$^{5}$  & \ldots         & 10    &$-$1.32  & 1.47        & 0.20  \\  
Yb & 70 &0.85     & $-$0.41$^{5}$                               &\ldots          & 2     &$-$0.70  & 1.35        & 0.21  \\  
Lu & 71 &0.10     & \ldots                                      &\ldots          & 2     &$-$1.14  & 1.66        & 0.08  \\  
Hf & 72 &0.85     & $-$0.59$^{1}$, $-$0.73$^{4}$,$-$0.72$^{5}$  & \ldots         & 6     &$-$0.88  & 1.17        & 0.09  \\  
Ta & 73 &$-$0.15\pp  & \ldots                                   & $-$1.60$^{4}$  &\ldots &\ldots   & 1.45   &\ldots        \\  
W  & 74 &0.79     & \ldots                                      & $-$0.90$^{4}$  &\ldots &\ldots   & 1.21   &\ldots        \\  
Re & 75 &0.26     & \ldots                                      & $-$0.21$^{4}$  &\ldots &\ldots   & 2.43   &\ldots        \\  
Os & 76 &1.35     & +0.43$^{1}$                                 & $-$0.07$^{4}$  & 3     & \pp0.23     & 1.78   & 0.22        \\  
Ir & 77 &1.32     & +0.20$^{1}$                                 & \,+0.18$^{4}$    & 5     & \pp0.10     & 1.68   & 0.18        \\  
Pt & 78 &1.61     & \ldots                                      & \,+0.30$^{3}$    & 2     & \pp0.00     & 1.29   & 0.30        \\  
Au & 79 &0.91     & \ldots                                      & $-$1.00$^{3}$  & 1     &$-$1.39  & 0.60   & \,~0.16$^{2}$   \\  
Pb & 82 &1.95     & $-$0.55$^{2}$                               & $-$0.65$^{3}$  & 1     &$-$0.65  & 0.30   &\ldots       \\  
Bi & 83 &0.65     & \ldots                                      & $-$0.40$^{3}$  & 1     &  $-$0.20 & 2.05  & 0.28        \\  
Th & 90 &0.03     & $-$0.98$^{1}$                               & \ldots         & 6     &$-$1.04  & 1.83   & 0.10        \\  
U  & 92 &$-$0.54\pp  & $-$1.92$^{1}$                            & \ldots         & 1     &$-$1.92  & 1.52   & 0.14        \\  

\hline
\hline        
\end{tabular}
\end{table*}

\setlength{\tabcolsep}{2pt}
\begin{table}
\centering 
\scalefont{0.88}
\caption{Abundances for CS~31082-001 and their uncertainties from analysis of: a) STIS spectra plus UVES smoothed spectra (tagged as SU, where those in bold face were derived using STIS), b) Keck-HIRES spectra (tagged as Keck), c) non-convolved UVES spectra (tagged as UVES). The stellar parameter uncertainty ($\sigma_{\rm param}$) was adopted from Hill et al. (2002), Sneden et al. (2009) and Siqueira-Mello et al. (2013), except for those with uncertainties presented in Table \ref{errors} (Ge, Mo, Sn, Ho, Yb, Lu, Hf, Pt, Bi). 
The statistical (fitting) uncertainties in cols.~6, 8 and 10 are internal errors, or in the case of a unique line, is an estimation of the fitting error.}\label{heavy_error}                   
\begin{tabular}{lccccccccc} 
\hline\hline             
Element  & Z &  $\sigma_{\rm param}$  & N & A(X)$_{\rm SU}$ & $\sigma_{\rm SU}$ & A(X)$_{\rm Keck}$ & $\sigma_{\rm Keck}$ & A(X)$_{\rm UVES}$ & $\sigma_{\rm UVES}$   \\
\hline  
\hline     
Ge & 32 & 0.150 & 1     &  \pp0.48  &    0.05    &\ldots  &\ldots & \pp0.48   &\ldots \\
Sr & 38 & 0.092 & 2     &  \pp0.40  &    0.02    &  \pp0.35  &  0.00 & \pp0.40   &0.00   \\
Y  & 39 & 0.121 & 22    &  $-$0.21  &    0.06    & $-$0.48  &  0.14 & $-$0.37  &0.17   \\
Zr & 40 & 0.114 & 66    &  \pp0.36  &    0.10    & \pp0.27   &  0.14 & \pp0.26   &0.15   \\
Nb & 41 & 0.128 & 3     &  $-$0.55  &    0.00    & $-$0.70  &  0.05 & $-$0.72  &0.05   \\
Mo & 42 & 0.070 & 1     &  $-$0.11  &  \ldots    & $-$0.24  &\ldots & $-$0.21  &0.05   \\
Ru & 44 & 0.159 & 5     &  \pp0.27  &    0.04    & \pp0.18   &  0.02 & \pp0.18   &0.02   \\
Rh & 45 & 0.162 & 3     &  $-$0.42  &    0.00    & $-$0.49  &  0.04 & $-$0.51  &0.09   \\
Pd & 46 & 0.166 & 5     &  $-$0.10  &    0.06    & $-$0.17  &  0.10 & $-$0.21  &0.07   \\
Ag & 47 & 0.166 & 2     &  $-$0.96  &    0.05    & $-$0.94  &  0.02 & $-$0.94  &0.07   \\
Sn & 50 & 0.200  & 1    &  $-$0.90  &  \ldots    &\ldots  &\ldots & $<$$-$0.40  &0.10   \\
Ba & 56 & 0.122 & 1     &  \pp0.15  &  \ldots    & \pp0.00   &\ldots & $-$0.05  &0.02   \\
La & 57 & 0.128 & 6     &  $-$0.65  &    0.08    & $-$0.72  &  0.11 & $-$0.73  &0.07   \\
Ce & 58 & 0.119 & 18    &  $-$0.31  &    0.00    & $-$0.40  &  0.09 & $-$0.41  &0.11   \\
Pr & 59 & 0.119 & 3     &  $-$0.86  &    0.00    & $-$0.94  &  0.02 & $-$0.94  &0.02   \\
Nd & 60 & 0.119 & 24    &  $-$0.13  &    0.00    & $-$0.34  &  0.10 & $-$0.33  &0.09   \\
Sm & 62 & 0.122 & 26    &  $-$0.48  &    0.12    & $-$0.56  &  0.17 & $-$0.54  &0.18   \\
Eu & 63 & 0.120 & 6     &  $-$0.78  &    0.04    & $-$0.96  &  0.10 & $-$0.93  &0.07   \\
Gd & 64 & 0.120 & 39    &  $-$0.28  &    0.09    & $-$0.44  &  0.13 & $-$0.45  &0.10   \\ 
Tb & 65 & 0.121 & 11    &  $-$1.26  &    0.00    & $-$1.23  &  0.09 & $-$1.22  &0.08   \\
Dy & 66 & 0.118 & 28    &  $-$0.22  &    0.05    & $-$0.28  &  0.09 & $-$0.25  &0.08   \\
Ho & 67 & 0.140 & 3     &  $-$0.97  &    0.05    & $-$0.98  &  0.06 & $-$0.98  &0.06   \\
Er & 68 & 0.142 & 21    &  $-$0.29  &    0.04    & $-$0.39  &  0.09 & $-$0.38  &0.10   \\
Tm & 69 & 0.115 & 10    &  $-$1.22  &    0.12    & $-$1.25  &  0.16 & $-$1.32  &0.16   \\
Yb & 70 & 0.210 & 2     &  $-$0.70  &   0.02     & $-$0.62  &  0.22 & $-$0.67  & 0.00 \\
Lu & 71 & 0.060 & 2     &  $-$1.14  &    0.00    & $-$0.94  &  0.00 & $-$1.14  &0.05   \\
Hf & 72 & 0.070 & 6     &  $-$0.64  &    0.11    & $-$0.85  &  0.06 & $-$0.88  &0.05   \\
Os & 76 & 0.157 & 3     &  \pp0.20   &   0.07     & \pp0.18   &  0.08 & \pp0.23   &0.16   \\
Ir & 77 & 0.140 & 5     &  \pp0.22   &   0.16     & \pp0.20   &  0.14 & \pp0.10   &0.12   \\
Pt & 78 & 0.040 & 4     &  $-$0.80   &    0.30    & $-$1.10   &  0.00 & $-$0.80 & 0.30  \\
Bi & 83 & 0.200 & 1     &  $-$0.20  &    0.00    &\ldots  &\ldots &\ldots  &\ldots \\
Th & 90 & 0.132 & 6     &  $-$0.98  &    0.00    & $-$0.96  &  0.09 & $-$1.04  &0.08   \\
\hline
\hline        
\end{tabular}
\end{table}

\begin{table}
\caption{Abundance uncertainties from stellar parameters for CS~31082-001,
 for uncertainties of $\Delta$T$_{\rm eff}$ = 100 K,
$\Delta$log g = 0.3, $\Delta$v$_{\rm t}$ = 0.2 km s$^{-1}$ 
for the elements not studied byy Hill et al. (2002).}\label{errors}
\begin{flushleft}
\small
\tabcolsep 0.15cm
\centering
\begin{tabular}{lccccc@{}c@{}}
\noalign{\smallskip}
\hline
\noalign{\smallskip}
\hline
\noalign{\smallskip}
\hbox{Element} & \hbox{$\Delta$T$_{\rm eff}$} & \hbox{$\Delta$log $g$} & 
\phantom{-}\hbox{$\Delta$v$_{\rm t}$} & \phantom{-}\hbox{($\sum$x$^{2}$)$^{1/2}$}&
 \\
\hbox{} & \hbox{100 K} & \hbox{0.3 dex} & \hbox{0.2 kms$^{-1}$} & & & \\
\hbox{(1)} & \hbox{(2)} & \hbox{(3)} & \hbox{(4)} & \hbox{(5)}  &\\
\noalign{\smallskip}
\hline
\noalign{\smallskip}
\noalign{\hrule\vskip 0.1cm}
\hbox{[Ge/Fe]}      &  +0.15  & +0.03      & +0.00     &  0.15   &\\
\hbox{[Mo/Fe]}      &  +0.01  & +0.06      &+0.03      &  0.07   &\\ 
\hbox{[Sn/Fe]}      &  +0.20  & +0.00      &+0.00      &  0.20   &\\ 
\hbox{[Ho/Fe]}      &  +0.02  & $\pm$0.10  &$\pm$0.10  &  0.14   &\\ 
\hbox{[Yb/Fe]}      &  +0.06  & +0.03      &$-$0.20      &  0.21   &\\ 
\hbox{[Lu/Fe]}      &  +0.01  & +0.06      &+0.00      &  0.06   &\\ 
\hbox{[Hf/Fe]}      &  +0.03  & +0.06      &+0.00      &  0.07   &\\ 
\hbox{[Pt/Fe]}      &  +0.03  & $-$0.03      &+0.00      &  0.04   &\\ 
\hbox{[Bi/Fe]}      &  +0.20  & $-$0.02      &+0.00      &  0.20   &\\ 
\noalign{\smallskip} 
\hline 
\end{tabular}
\end{flushleft}
 \end{table}

The presence of heavy elements in very old metal-poor halo stars should be due to
a source of r-process elements in the early Galaxy, as suggested by Truran (1981), 
given that there might not be enough time for AGB stars to form these elements in an s-process.
On the other hand, Frischknecht et al. (2012, 2016), and Limongi \& Chieffi (2018) demonstrated that metal-poor rotating massive stars are able to 
synthesize neutron-capture elements through an s-process.

Sneden et al. (2008, and references therein) reviewed the formation of heavy elements through the s- and r- processes and showed that the abundance pattern of the heaviest r-process elements, i.e., those with atomic number 56 $<$ Z $<$ 78 is the same in different objects, including the Sun.
The same does not apply to the trans-iron elements (31 $<$ Z $<$ 52) and the actinides (89 $<$ Z $<$ 92).

In general terms, the production of the lightest neutron-capture elements (that appear to vary from star-to-star) has been assigned to a Light Element Primary Process (LEPP; Travaglio et al. 2004), an i-process (Cowan \& Rose 1977; Roederer et al. 2016), neutrino-driven winds in supernovae (Arcones \& Montes 2011, and references therein), or a weak r-process (Wanajo 2013); see further discussion by Spite et al. (2018) and Peterson et al. (2020).

The r-process was more recently reviewed by Cowan et al. (2021). Traces of r-process element 
production was confirmed from the neutron-star merger (NSM) kilonova AT2017gfo, that was associated with the gravitational-wave event GW170817 (Tanvir et al.  
2017; Drout et al. 2017). Other likely sources are mergers of neutron stars with a black hole, 
magneto-rotational supernovae (Winteler et al. 2012), and collapsars (Siegel et al. 2019).

The nucleosynthesis of neutron-capture elements in massive stars has been explored by several studies, including Wanajo (2007), Just et al. (2015), Banerjee et al. (2017), and Ritter et al. (2018). Here we compare the r-process abundances for CS~31082-001 with calculations from Wanajo (2007) and Just et al. (2015).

For background, the neutrino winds that arise in core-collapse supernovae are expected to produce all of the r-process elements until the actinides. Wanajo et al. (2002) computed a model with a compact neutron star of 2.0 M$_{\odot}$, and a  neutrino sphere of 10 km. They established a freezeout temperature, specified as the final temperature of the neutrino winds, to be T$_{9f}$\,$=$\,1.0 (in units of 10$^9$\,K), to better reproduce the solar r-process pattern near the third peak.
Wanajo (2007) adopted T$_{9f}$\,$=$\,0.1 instead, which he called the cold r-process, in which the neutron capture competes with the $\beta$-decay, similar to the hot r-process. The cold r-process also predicts a low lead production, that is compatible with its estimated abundance in CS~31082-001. 

In Fig.~\ref{wanajo} we show the abundance pattern of CS~31082-001 compared with the solar pattern (Asplund et al. 2021), and the models of a neutrino wind scenario for a hot r-process (Wanajo et al. 2002), and for a cold r-process (Wanajo, 2007). The lighter neutron-capture elements are normalized to the Zr abundance, and the heavier elements are normalized to the Eu abundance (see discussion below).

Wanajo et al. (2014) proposed the first yield calculations of a NSM, based on full general-relativistic, approximate neutrino transport simulations for masses of 1.3 to 1.35\,M$_{\odot}$. These appeared as a promising candidate for r-process enrichment in the Galaxy, given that the ejecta consist of almost pure neutrons. Models from Just et al. (2015) also included the black hole torus ejecta, formed after the NSM. 

In Fig.~\ref{just} the abundances of CS~31082-001 are compared with the models from Just et al. (2015) for torus masses of 0.03, 0.1 and 0.3\,M$_{\odot}$. We see a general agreement, with Zr enhanced, Pd-Sn deficient, and Hf-Pt enhanced. At least for the abundance pattern, there is reasonable agreement with the NSM models.

Adopting the nuclesynthesis predictions from Wanajo et al. (2014) and Just et al. (2015), Kobayashi et al. (2022) revised the r-process enrichment, intending to investigate if NSMs could be the main contributor to r-process elements in the Galaxy. Their models adopting magneto-rotational supernovae (adopting the calculations from Nishimura et al. 2015) best reproduce the observations, so they concluded that NSMs are probably not the main contributor to r-process elements. 
Alongside this, we note that Holmbeck et al. (2019) found that the production of actinides
is probably due to NSMs, and that they also concluded that another source was required to account for the observed r-process abundances.


\subsection{Trans-iron elements: Ge, Sr, Y, Zr, Nb, Mo, Ru, Rh, Pd, Ag, Sn}

The so-called trans-iron elements, with atomic numbers 31\,$<$\,Z\,$<$\,52, include elements from Ga (Z\,$=$\,31) to Te (Z\,$=$\,52). 

Germanium can be considered the last of the upper iron-peak group or the first of the neutron-capture elements (Woosley \& Weaver 1995), or both at the same time (Niu et al. 2014).
To further understand the origin of Ge, Cowan et al. (2005)
and Siqueira-Mello et al. (2013) compared its abundance in 
metal-poor stars with that of the r-process element Eu. They showed that
the Ge abundance correlates with metallicity, but not with Eu, suggesting that Ge is not a typical
r-process element.

As discussed above, and pointed out by Kratz et al. (2007),  
the trans-iron element abundances
vary from star to star and require an additional
secondary weak r-process site (Spite et al. 2018; Peterson et al. 2020).
Roederer et al. (2022a,b) instead suggested that the universality of the r-process
accepted in the literature for the 56\,$<$\,Z\,$<$\,71 from Ba to Lu, and the third peak
r-process elements with 72\,$<$\,Z\,$<$\,78 from Hf to Pt, might also apply to first peak s-elements, or light-s elements (ls) Se, Sr, Y, Zr, and Mo, provided that these are scaled
independently from the other elements.

\begin{figure*}
    \centering
    \includegraphics[width=5.5in]{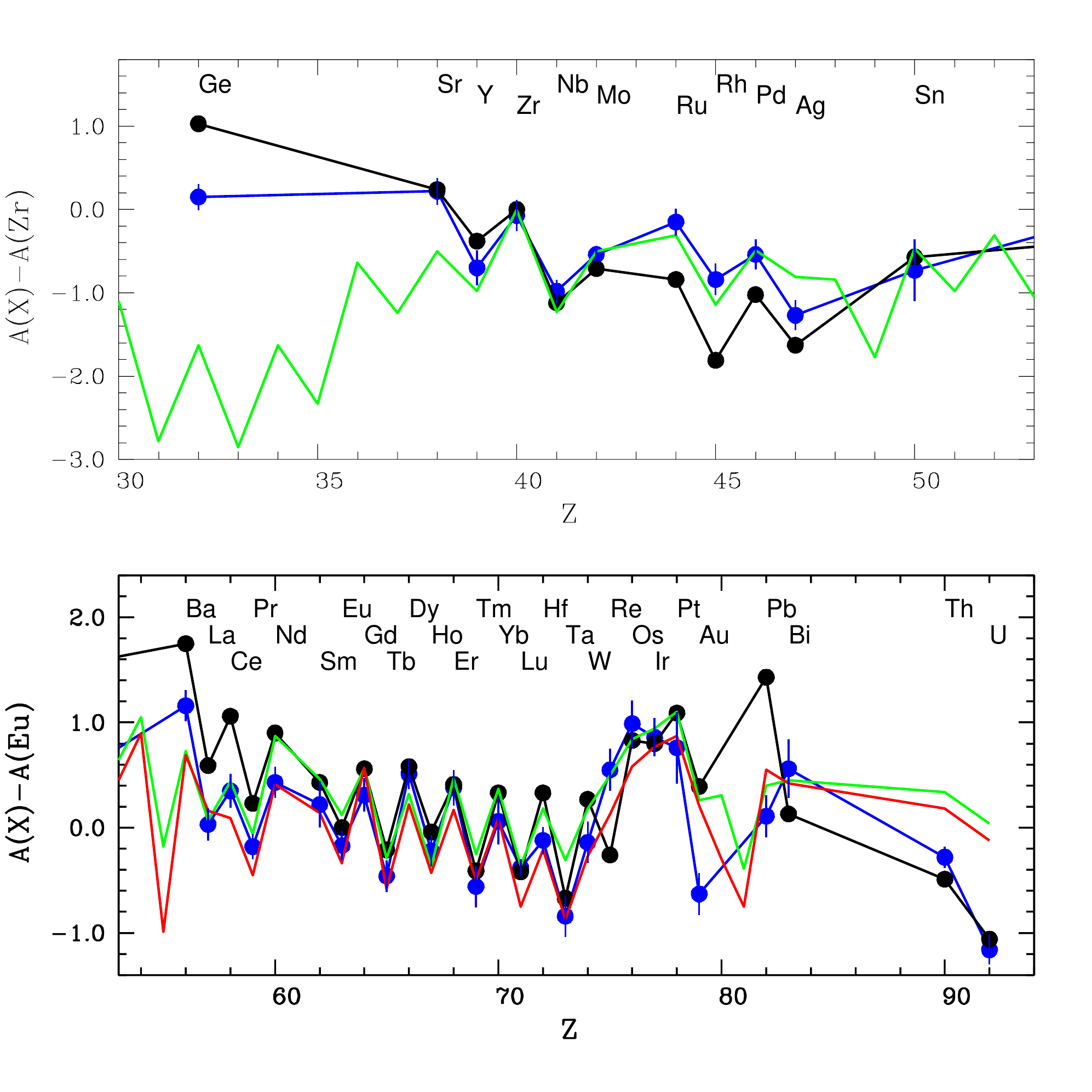}
    \caption{Abundance patterns of CS~31082-001 (blue) and the Sun (black) compared with models for a neutrino wind scenario for a hot r-process (red, from Wanajo et al. 2002) and for a cold r-process (green, from Wanajo, 2007). Upper panel: lower neutron-capture elements from Ge (Z\,$=$\,32) to Sn (Z\,$=$\,50), normalized to the abundance of Zr. Lower panel: heavier neutron-capture elements from Ba (Z\,$=$\,56) to U (Z\,$=$\,92), normalized to the abundance of Eu. We have assumed uncertainties of $\pm$0.20 for Ta, W, Re, Au, Pb.}\label{wanajo}
\end{figure*}
\begin{figure*}
    \centering
    \includegraphics[width=6.5in]{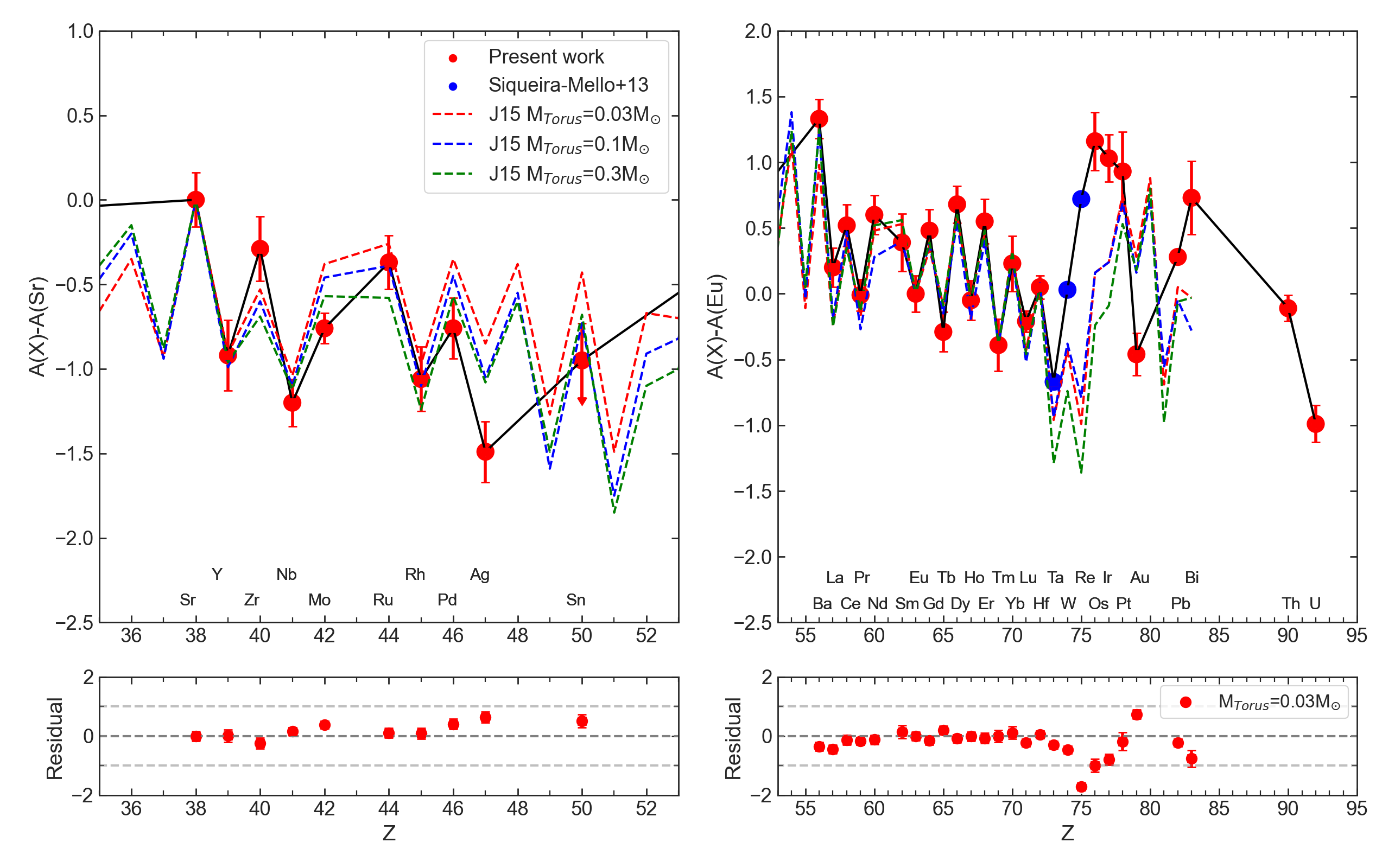}
    \caption{Abundance pattern of CS~31082-001 (red and blue dots linked by the black line) compared with models from Just et al. (2015) for black-hole torus masses of 0.03, 0.1, and 0.3 M$_{\odot}$. Left: abundances are normalized to Sr showing the elements from Sr to Sn. Right: elements from Ba to U with abundances normalized to Eu. The lower panels show the residuals between the model with M$_{Torus}$\,$=$\,0.03 M$_{\odot}$ and the abundance pattern of CS~31082-001.}\label{just}
\end{figure*}

From Fig.~\ref{wanajo} it appears that, for the lighter neutron-capture elements, the abundance pattern of the trans-iron elements Sr to Mo is rather similar to solar, and Mo to Ag are enhanced, 
whereas Sn is deficient in CS~31082-001. This is in agreement with the suggestion from Roederer et al. (2022a,b) that the Sr to Mo abundance pattern is universal, whereas from Ru through to Sn it does not appear so. The hot and cold r-process models from Wanajo (2007) coincide in this element number range (38\,$<$\,Z\,$<$\,42), and fit the abundances of CS~31082-001 well (as previously discussed by Barbuy et al. 2011; Siqueira-Mello et al. 2013).

\subsection{Second peak elements: Ba, La, Ce, Pr, Nd}

Ba, La, Ce, Pr, and Nd correspond to second peak s-elements or heavy-s (hs).
These elements are dominantly s-process elements in stars like the Sun, but
in CS~31082-001 their production is likely due to an early r-process.
They show a mean enhancement of [(Ba, La, Ce)/Fe]\,$=$\,$+$1.0 and  [(Pr, Nd)/Fe]\,$=$\,$+$1.18
(see Table \ref{simmerer}).

These elements are well fit by the hot and cold models from Wanajo (2007), but they are deficient relative to the solar abundance pattern (Fig.~\ref{wanajo}).

\subsection{Very enhanced r-process elements: Sm, Eu, Gd, Tb, Dy, Ho, Er, Tm, Yb, Lu, Hf, Os, Ir, Pt,  Au}

The elements discussed in this subsection are dominantly r-process elements in the Sun, and they are strongly enhanced in CS~31082-001.

The elements from Sm (Z\,$=$\,62) to Lu (Z\,$=$\,71), show a mean r-process enhancement
of [r-elements/Fe]\,$=$\,+1.45. These include the first estimates for Ho and Yb from these data, which are close to the mean with [Ho/Fe]\,$=$\,$+$1.44 and [Yb/Fe]\,$=$\,$+$1.35 (see Table \ref{simmerer}). Os (Z\,$=$\,76) and Ir (Z\,$=$\,77) also show a mean r-process enhancement of [r-elements/Fe]\,$=$\,$+$1.73, and the actinides Th (Z\,$=$\,90) and U (Z\,$=$\,92) show [r-elements/Fe]\,$=$\,$+$1.68.

Hf (Z\,$=$\,72) is produced as a r- and s-element in proportions of 0.51 and 0.49, respectively, in the solar mixture (Simmerer et al. 2004,) and can potentially discriminate the site production of r-process elements (Eichler et al. 2019). The Hf/Eu and Hf/Th ratios show the balance between the NSMs that mostly produce Hf and the type II supernovae responsible for Eu and Th. The log $\epsilon$(Th/Hf)\,$=$\,A(Th)\,$-$\,A(Hf)\,$=$\,$-$0.16 in CS~31082-001 is compatible with the r-process production from Eichler et al. (2019; lower-left panel of their Fig.~2). The same conclusion is reached from the value of A(Hf)\,$-$\,A(Eu)\,$=$\,$+$0.05 in CS~31082-001, that corresponds to a pure r-process enrichment of Hf (see Roederer et al. 2009, upper-left panel of their Fig.~6).

The estimate for Au (Z\,$=$\,79) from one STIS line, revised with respect to Barbuy et al. (2011), results in a lower enhancement of [Au/Fe]\,$=$\,$+$0.60; in principle, Au is almost entirely due to the r-process.

The heavier neutron-capture r-process elements are well fit by the solar abundance pattern as well as by the hot and cold models from Wanajo (2007), as shown in Fig.~\ref{wanajo}. This pattern can be considered as universal as explained by Roederer et al. (2022a,b and references therein).

\subsection{Heaviest stable elements: Pb, Bi}

Pb and Bi could be considered as a third peak of the s-elements (Sneden et al. 2008),
or early production as a r-product, as discussed by Plez et al. (2004).
Cowan et al. (1999) reported that calculations reproduce the solar 
isotopic r-abundances, including the heaviest stable Pb and Bi isotopes, at the same time that
about 85\% of Pb and Bi are formed through the radioactive decay of Th and U.

The solar isotopic fractions from Asplund et al. (2009) were 1.997, 18.582, 20.563 and 58.858 for $^{204}$Pb, $^{206}$Pb, $^{207}$Pb and $^{208}$Pb, respectively. Plez et al. (2004) investigated what would be the minimum amount of lead due to the decay of $^{238}$U and $^{232}$Th into $^{206}$Pb, and $^{208}$Pb, respectively, for the Sun. They then also estimated this for CS~31082-001 by adopting an age of 13.5\,$\pm$\,1.5~Gyr and decay times of $\tau$\,$=$\,4.47 and 14.05~Gyr for $^{238}$U and $^{232}$Th, respectively. The conversion $^{235}$U into $^{207}$Pb was also computed but using theoretical values. The resulting lead abundance of $-$0.61\,$<$\,A(Pb)\,$<$\,$-$0.55, was very close to the observed Pb abundance, with the conclusion that effectively
all of the Pb was a result of the decay of $^{238}$U, $^{232}$Th and $^{235}$U.

As discussed by Plez et al. (2004) and Barbuy et al. (2011), nucleosynthesis calculations predict that the total Pb abundance is
the sum of any assumed initial r-process Pb production, plus the radioactive decay products of Th and U, and suggest a higher Pb abundance for CS~31082-001 than observed. According to Roederer et al. (2009), stars with A(La)\,$-$\,A(Eu)\,$>$\,$+$0.25 show some amount of s-process material, whereas those with $+$0.09\,$<$\,A(La)\,$-$\,A(Eu)\,$<$\,$+$0.23 show a pure r-process content. In CS~31082-001 this quantity gives A(La)\,$-$\,A(Eu)\,$=$\,$+$0.20, typical of r-process nucleosynthesis. Therefore, it should not show any s-process contribution to Pb, and its value of A(Pb)\,$-$\,A(La)\,$=$\,$+$0.08 is compatible, and even lower,  than that from a pure r-process (Roederer et al. 2009, see lower panel of their Fig.~5).

For Bi, the s-process terminates at $^{209}$Bi, the last stable isotope, which is actually radioactive but with a half-life longer than a Hubble time. Its abundance in CS~31081-001 is compatible with r-process production (Fig.~\ref{wanajo}) but appears enhanced relative to the predictions from Just et al. (2015, see Fig.~\ref{just}).
The result of [Bi/Fe]\,$=$\,$+$2.05 shows a r-process enhancement comparable
to the other elements studied here. However, if we adopt the oscillator strength indicated in NIST
of log~gf\,$=$\,$-$0.15, instead of the log~gf\,$=$\,1.35 previously available and adopted here, we would
have [Bi/Fe]\,$=$\,$+$3.75, which is unexpectedly high. It is difficult to understand why Bi would show a s-process contribution in this star, while Pb does not, but it is also difficult to accept that this element is much more enriched in some r-process nucleosynthesis than any of the other elements. Even if s-process heavy-element production by spinstars 
is included, Bi is enhanced but not more than the other studied elements (Frischknecht et al. 2016).

In short, results for both Pb and Bi are intriguing and remain open for further studies.

\subsection{Actinides: Th, U}

The Th and U abundances for CS~31082-001 were discussed extensively by Cayrel et al. (2001) and Hill et al. (2002), and its age by Barbuy et al. (2011), so we do not revisit these points here.
Th and U are entirely r-process element, and their enhancements of [Th/Fe]\,$=$\,$+$1.83 and [U/Fe]\,$=$\,$+$1.52 indicate they are produced in the same process as the other r-process elements.

\bigskip
To summarize this section, following the proportion of r- to s-process contributions presented by Simmerer et al. (2004) and Prantzos et al. (2020) for the solar mixture, the mean abundance enhancements from the r-process- or s-process-dominated elements for different groups are given in Table~\ref{simmerer}. It is clear that the r-process dominated elements are more enhanced than others, as seen in Fig.~\ref{fig1}. This figure also compares the r-process abundance pattern of CS~31082-001 to that of the Sun, that we obtained by multiplying each element abundance adopted in the star with the solar r-fraction  from Prantzos et al. (2020), normalized to the abundance of Au; this pattern
again shows the r-process nature of the neutron-capture elements in CS~31082-001.

\section{Conclusions}

The halo star CS31082-001 is among the most completely analysed in terms of elemental abundances, with estimates for 60 elements. Past results were available for 54 elements from Hill et al. (2002), Plez et al. (2004), Barbuy et al. (2011) and Siqueira-Mello et al. (2013), which we have supplemented with abundances for Be, V and Cu from E22, and now Sn, Ho, and Yb in this study. The only star with a greater number of elemental abundances is HD~222925, with estimates for 63 elements (Roederer et al. 2022b). However, with [Fe/H]\,$=$\,$-$1.4, the metallicity of HD~222925 is very different to that of CS~31082-001. This means that HD~222925 was probably enriched later, and included s-process contributions from AGB stars, whereas the heavy-element abundances in CS~31082-001 are all due to the r-process (except for the possibility of s-process elements from spinstars).\smallskip

\begin{figure*}
    \centering
    \includegraphics[width=7.0in]{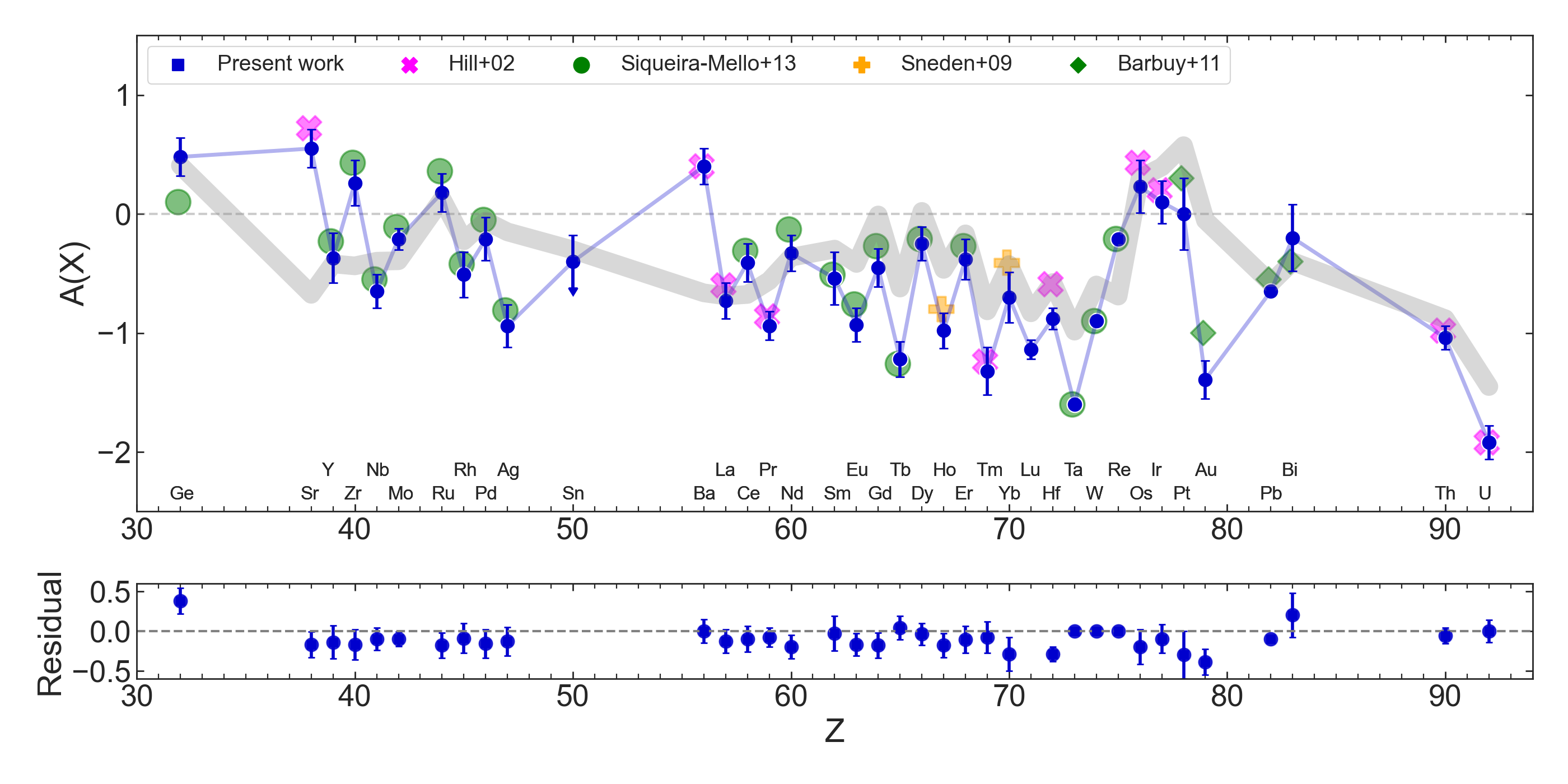}
    \caption{Abundance pattern of CS~31082-001 comparing the non-normalised abundances obtained in the present work using the near-UV lines with those from the visible (Hill et al., 2002) and UV regions (Sneden et al., 2009; Barbuy et al. 2011; Siqueira-Mello et al. 2013). The residuals are shown in the lower panel from the abundance comparison between the present work and the literature. The shaded grey line is the solar pattern corresponding to the element abundance multiplied by its r-fraction normalised to the abundance of Au.}\label{fig1}
\end{figure*}

We now briefly summarise our main conclusions:

1) As a general conclusion, we confirm the r-process dominated elements are very enhanced and compatible with the r-process pattern of the Sun (Table \ref{simmerer} and Fig.~\ref{fig1}).

\setlength{\tabcolsep}{10pt}
\begin{table*}
\scalefont{0.88}
\caption{r- and s-process contributions from Simmerer et al. (2004) and Prantzos et al. (2020) for solar composition and abundance enhancements of different groups of elements. The fractions by Prantzos et al. (2020) that do not add to 100\% are due to
a contribution from the p-process. ":" means uncertain.
}\label{simmerer}  
\centering                  
\begin{tabular}{lcccccccc} 
\hline\hline             
Element  &  Z  & r-fraction & s-fraction & A(X)$_{\rm adopted}$ & [X/Fe]$_{\rm adopted}$ & [X/Fe]$_{\rm Mean}$ \\
\hline  
\hline
Ge & 32   & 0.431/0.364  & 0.569/0.636   & \pp0.48      &$-$0.24\pp &  $-$0.24 \\
\hline\hline
Sr & 38   & 0.11~\,/0.083   & 0.89~\,/0.912     & \pp0.55      &0.62   & \\
Y  & 39   & 0.281/0.222  & 0.719/0.778    & $-$0.37   &0.32    &  +0.50\\
Zr & 40   & 0.191/0.183  & 0.809/0.817    & \pp0.26      &0.57    & \\  
\hline\hline
Nb & 41   & 0.324/0.349   & 0.676/0.651    & $-$0.65   &0.78    & +0.80 \\  
Mo & 42  &  0.323/0.275   & 0.677/0.497    & $-$0.21   &0.81    & \\ 
\hline\hline
Ru  & 44  & 0.61~\,/0.591   & 0.39~\,/0.338     & \pp0.18      &1.33    & \\ 
Rh  & 45  & 0.839/0.878  & 0.161/0.122    & $-$0.51   &1.61    &  +1.27 \\ 
Pd  & 46  & 0.555/0.542  & 0.445/0.448    & $-$0.21   &1.12    & \\ 
Ag  & 47  & 0.788/0.791  & 0.212/0.209    & $-$0.94   &1.00    & \\ 
\hline\hline
Sn  & 50  & 0.225/0.301  & 0.775/0.680    & $-$0.40      &0.48    &  +0.48\\
\hline
\hline
Ba  & 56  & 0.147/0.109  & 0.853/0.888    & \pp0.40      &1.03    & \\  
La & 57   & 0.246/0.200  & 0.754/0.799    & $-$0.73   &1.06    &  +1.00 \\ 
Ce & 58   & 0.186/0.148  & 0.814/0.848    & $-$0.41   &0.91    & \\ 
\hline
\hline
Pr & 59   & 0.508/0.465  & 0.492/0.535    & $-$0.94   &1.21   &  +1.18\\ 
Nd & 60   & 0.421/0.385  & 0.579/0.615    & $-$0.133   &1.15    & \\
\hline
\hline
Sm & 62   & 0.669/0.647  & 0.331/0.325    & $-$0.54   &1.41     &\\ 
Eu & 63   & 0.973/0.951  & 0.027/0.049    & $-$0.93   &1.45     &\\ 
Gd & 64   & 0.819/0.835  & 0.181/0.163    & $-$0.45   &1.37     & \\  
Tb & 65   & 0.933/0.928  & 0.067/0.072    & $-$1.22   &1.37     & \\ 
Dy & 66   & 0.879/0.847  & 0.121/0.151    & $-$0.25   &1.55     & +1.45  \\   
Ho & 67   & 0.936/0.926  & 0.064/0.074    & $-$0.88   &1.44     & \\
Er & 68   & 0.832/0.799  & 0.168/0.184    & $-$0.38   &1.59     & \\ 
Tm & 69   & 0.829/0.872  & 0.171/0.128    & $-$1.32   &1.47     & \\ 
Yb & 70   & 0.682/0.570  & 0.318/0.429    & $-$0.70   &1.35     & \\
Lu & 71   & 0.796/0.796  & 0.204/0.204    & $-$1.14   &1.66     & \\
\hline
\hline
Hf & 72   & 0.51~\,/0.393   & 0.49 /0.605     & $-$0.88   &1.17     & \\ 
Ta  & 73  & 0.588/0.497   & 0.412/0.503    & $-$1.60   &1.45     &  +1.28 \\ 
W   & 74  & 0.462/0.397   & 0.538/0.601    & $-$0.90   &1.21     & \\ 
\hline
\hline
Re  & 75  & 0.911/0.846  & 0.089/0.154    & $-$0.21   &2.41     &  +2.41\\ 
\hline
\hline
Os  & 76  & 0.916/0.897  & 0.084/0.103    & \pp0.23      &1.78     & \\ 
Ir  & 77  & 0.988/0.989  & 0.012/0.011    & \pp0.10      &1.68     &  +1.73 \\ 
\hline
\hline
Pt  & 78  & 0.949/0.922  & 0.051/0.078    & \pp0.00   &1.29     & +1.29: \\ 
\hline
\hline
Au  & 79  & 0.944/0.942  & 0.056/0.058    & $-$1.39   &0.60     & +0.45 \\ 
Pb  & 82  & 0.214/0.169  & 0.786/0.831    & $-$0.65   &0.30     & \\
\hline
\hline
Bi  & 83  & 0.647/0.784  & 0.353/0.216    & $-$0.20   &2.05     &  +2.05 \\ 
\hline
\hline
Th & 90   & 1.00/1.00   & 0.00/0.00     & $-$1.04   &1.83     &  +1.68\\
U  & 92   & 1.00/1.00   & 0.00/0.00     & $-$1.92   &1.52     & \\
\hline
\hline                          
\end{tabular}
\end{table*}

2) Ge is deficient relative to the solar abundance pattern, indicating that Ge is dominantly
an iron-peak element and not a neutron-capture element (except perhaps for a small fraction).

3) Ho is enhanced by [Ho/Fe]\,$=$\,$+$1.44, which is compatible with the enhancements of
Gd, Tb, Dy, Tm and Lu, although somewhat lower than than those of a few other elements.
Roederer et al. (2022b) pointed out that for HD~222925, the Ho abundance is the most discrepant relative to the solar pattern,  but this is not the case for CS~31082-001.

4) Our estimate for Yb is compatible with the other r-element enhancements in CS~31082-001.

5) With [Sn/Fe]\,$=$\,0.48, the enhancement of Sn is compatible with that of trans-iron elements Mo and Ag, but is lower than seen for others (e.g. Ru, Rh, Pd). The first Sn detection in a r-process-rich star was for HD~222925 from Roederer et al. (2022b). Our estimate for CS~31082-001 is therefore the second, albeit only an upper limit.

6) Elements in the range 31 $\leq$ Z $\leq$ 50, i.e. from Ge to Sn, do not scale with
the solar abundance pattern (see upper panel of Fig.~\ref{wanajo}), thus agreeing with the conclusion of Roederer et al. (2022b).

7) Elements with Z $\geq$ 56 (i.e. Ba and heavier) match the solar abundance pattern (see lower panel of Fig.~\ref{wanajo}), as suggested previously by Cowan et al. (1999), Roederer et al. (2010) and Roederer et al. (2022b and references therein).

{8) The heavier elements Pt, Au, Pb and Bi are intriguing. The Pt abundance is very uncertain, and could be lower if the line computed with HFS was considered. The [Au/Fe] and [Pb/Fe] abundances appear too low relative to all other elements. For Bi we have conservatively adopted the oscillator strength of 1.35 previously available, but a value of $-$0.15 is now indicated in NIST, which would give a very high [Bi/Fe]\,$=$\,3.75, rather than the [Bi/Fe]\,$=$\,2.05 as adopted. Bi is the most intriguing element of all, and has not been well studied in the literature. Further studies of these elements in r-process-rich metal-poor stars should be pursued.

\smallskip
Finally, 
the characterization of the whole abundance pattern for a r-II star ([Eu/Fe]\,$>$\,$+$0.7, cf. Holmbeck et al. (2020), and [Ba/Eu]\,$<$\,0.0) can be a key aspect in determining not only the nucleosynthesis channels that create every element in the early stages of the Galaxy, but also the origin of these stars. Roederer et al. (2018) analysed the orbital parameters for r-II stars, including CS~31082-001, and suggested that stars with [Eu/Fe]\,$>$\,$+$0.7 are only found in halo-like orbits and were probably formed in a low star-formation efficiency environment similar to those found in dwarf galaxies. This finding is also supported by the comoslogical zoom-in simulations by Hirai et al. (2023), where the r-II stars are predominantly formed $\sim$10 Gyr ago within low-mass dwarf galaxies that were later disrupted. The time scale for the r-II stars from Hirai et al. (2023) is compatible with the age of 14.0\,$\pm$\,2.4~Gyr derived for CS~31082-001 by Hill et al. (2002) and Barbuy et al. (2011). Looking ahead, further study of the origins of the few known actinide-rich stars will be important to understand their physical properties and chemical-enrichment histories.

\section*{Acknowledgements}
HE acknowledges a CAPES PhD fellowship and a project grant from the Knut and Alice Wallenberg foundation (KAW 2020.0061 Galactic Time Machine). MC acknowledges a previous IC Fapesp fellowship 2020/14944-4 and a CNPq Master fellowship. BB acknowledges grants from FAPESP, CNPq and CAPES - Financial code 001.

Based on observations made with the ESO Very Large Telescope at Paranal Observatory,
Chile: Program ID 165.N-0276 (PI: R. Cayrel).
This research is also based on
observations made with the NASA/ESA {\it Hubble Space Telescope} obtained from the Space Telescope Science Institute, which is operated by the Association of Universities for Research in Astronomy, Inc., under NASA contract NAS 5–26555,
and with the NASA/ESA {\it Hubble Space Telescope (HST)} 
through the Space Telescope Science Institute, operated by the Association of 
Universities for Research in Astronomy Inc., under NASA contract NAS5-26555. These observations are associated with program 9359 (PI: R. Cayrel).
Some of the data presented herein were obtained at the W. M. Keck Observatory, which is operated as a scientific partnership among the California Institute of Technology, the University of California and the National Aeronautics and Space Administration. The Observatory was made possible by the generous financial support of the W. M. Keck Foundation. The Keck observations are related to program U53H (PI: M. Bolte).

\section*{Data Availability}

Observed data are available at the ESO and HST archives. Reduced data and
calculated synthetic spectra available under request.
 







\appendix

\section{Heavy element abundances}

Table~\ref{Alllines} presents the line list of the heavy elements studied in the
3000 to 4000 {\rm \AA} region. The abundances, log $\epsilon$(X)\,$=$\,A(X), are 
derived from the STIS or smoothed UVES spectra (column 4), the raw (unsmoothed) UVES spectra from the recent ESO reductions (column 5) and the Keck data (column 6); the final column of the table lists past references for each line.

\setlength{\tabcolsep}{5pt}
\begin{table} 
\caption{Abundances for CS~31082-001 for each near-UV, heavy-element line considered in this study.
The abudances were estimated from: a) STIS or UVES-smoothed spectra (col. 4), b) UVES-raw spectra (col. 5), c) Keck spectra (col. 6); abundances derived from the STIS spectra are in bold font. Sources for each line are indicated in the final column, as follows: 1: Hill et al. (2002); 2: Siqueira-Mello et al. (2013); 
3: Barbuy et al. (2011); 4: Cayrel et al. (2001);  5: Cowan et al. (2002);
  6: Sneden et al. (2003); 7: Roederer \& Lawler (2012); 8: Roederer et al. (2022b);
   9: Lawler et al. (2001a); 10: Lawler et al. (2001b); 11: Lawler et al. (2001c); 
   12: Lawler et al. (2001d); 13: Lawler et al. (2004);  14: Den Hartog et al. (2003); 
   15: Nilsson et al. (2002); 16: Nilsson et al. (2010); 17: Ljung et al. (2006); 
   18: Wiese \& Martin (1980) 19: Lawler et al. (2009): 20: Ivans et al. (2006); 21: Den Hartog et al. (2005).
   Symbols: $^a$ line not sensitive to element abundance. ":" uncertain; "::" very uncertain; "*" HFS taken into account. }
   \label{Alllines}
\centering   
\scalefont{0.90}
\begin{tabular}{lcccccccccccc}
\hline
\hbox{$\lambda$({\rm \AA})} & \hbox{$\chi_{\rm ex}$} & \hbox{log gf} &  &
{log$\epsilon$} & & Ref\\  
\cline{4-6}
&  &  & STIS/UVES & UVES$_{raw}$ & Keck &  & \\
\hline\hline    
    \midrule
    \multicolumn{7}{c}{{\ion{Ge}{I} (Z\,$=$\,32)}} \\
    \multicolumn{7}{c}{{log$\epsilon$(Ge) = A(Ge) = 0.48 }} \\
    \midrule
     3039.067 &0.883&  $-$0.040 & 0.48 & 0.48 & --- & 2 \\
     \shrinkheight{00pt}
    \midrule
    \multicolumn{7}{c}{{\ion{Sr}{II} (Z\,$=$\,38)}} \\
    \multicolumn{7}{c}{{log$\epsilon$(Sr) = A(Sr) = 0.55 - see text }} \\
    \midrule
    3464.453 &3.040&    0.530 & 0.37 &0.40 &0.40  & 6   \\
    3474.889 &3.040& $-$0.460 & 0.42 &0.40 &0.35: & 18   \\

    \shrinkheight{00pt}
    \midrule
    \multicolumn{7}{c}{{\ion{Y}{II} (Z\,$=$\,39)}} \\
    \multicolumn{7}{c}{{log$\epsilon$(Y) = A(Y)  = $-$0.37 }} \\
    \midrule
     3095.872 & 0.130 & $-$1.740 & $-$0.03&$-$0.03&-- & 8 \\   
     3135.168 & 0.180 & $-$1.680 & $-$0.03&$-$0.03&--  & 8 \\   
     3200.272 &0.130& $-$0.430 & $-$0.23&-0.43&--  & 2 \\
     3203.322 &0.104& $-$0.370 & $-$0.23&-0.23&-0.33  & 2 \\
     3216.682 &0.130& $-$0.020 & $-$0.23&-0.43&-0.63  & 2 \\
     3242.280  &0.180& 0.210 & $-$0.23&-0.23&-0.43 & 2,6 \\
     3327.878 &0.409& 0.130   & $-$0.23&-0.43&-0.53 & 6 \\
     3448.808 &0.409& $-$1.440 & $-$0.23&-0.13&-0.23  & 2 \\
     3549.005  &0.130& $-$0.28  & $-$0.23&-0.38&-0.43 & 2,5,6 \\
     3584.518  &0.104& $-$0.410 & $-$0.23&-0.23&-0.43 & 2 \\
     3600.741  &0.180& 0.280  & $-$0.23&-0.43&-0.43 & 2,5,6 \\
     3601.919  &0.104& $-$0.180 & $-$0.23&-0.43&-0.43 & 2  \\
     3611.044  &0.130& 0.110  & $-$0.23&-0.63&-0.53 & 2,5,6  \\
     3628.705 &0.130& $-$0.710 &  $-$0.23&-0.33&-0.33  & 2  \\
     3633.122 &0.000& $-$0.100 & $-$0.20&-0.63&-0.78  & 2  \\
     3710.294 &0.180& 0.460  & $-$0.23&-0.53&-0.73  & 2  \\
     3747.550 &0.104& $-$0.910 & $-$0.23&-0.43&-0.43  & 5,6 \\
     3774.331  &0.130& 0.210 & $-$0.23&-0.53&-0.53 & 1,2,5,6 \\
     3788.694  &0.104& $-$0.070 & $-$0.23&-0.43&-0.53 & 1,2,5,6 \\
     3818.341  &0.130& $-$0.980 & $-$0.26&-0.23&-0.23  & 1,2,6 \\
     3832.899  &0.180& $-$0.340 & $-$0.23&-0.53&-0.63 & 6 \\
     3950.352  &0.104& $-$0.490 & $-$0.23&-0.43&-0.53 & 1,2,5,6 \\
    \midrule
    \multicolumn{7}{c}{{\ion{Zr}{II} (Z\,$=$\,40)}} \\
    \multicolumn{7}{c}{{log$\epsilon$(Zr) = A(Zr) = 0.33 }} \\
    \midrule
    3019.832 &0.039&$-$1.130&  0.43&--&-- & 2 \\
    3028.045 &0.972&0.020&  0.33 &--&--  & 2 \\
    3030.915 &0.000&$-$1.040&  0.43 &0.33&--  & 2 \\
    3036.390  &0.559& $-$0.420 &  0.23 &0.23&--  & 5  \\
    3036.514 &0.527& $-$0.600 &  0.23 &0.23&--  & 5  \\
    3054.847 &0.713& $-$1.200 &  0.43 &0.43&--  & 2,5,6 \\
    3060.111 &0.039& $-$1.370 &  0.43 &0.03&--  & 5,6 \\
    3061.334 &0.095& $-$1.380 &  0.43:&0.33&--  & 2,5,6 \\
    3095.073  &0.039& $-$0.960 & 0.43&0.63&--  & 2 \\
    3125.926  &0.000& $-$0.883 & 0.43&0.43&--  & 2 \\
    3129.763  &0.527& $-$0.320 & 0.43&0.38&--  & 2 \\
    3133.489 &0.959&$-$0.200& 0.33&0.03&--  & 2 \\
    3138.683 &0.095&$-$0.460& 0.43&0.33&--  & 2 \\
    3231.692 &0.039&$-$0.590& 0.43&0.38&0.63 & 2 \\
    3241.042 &0.039&$-$0.504& 0.23&0.10&0.38 & 2 \\
    3272.221 &0.000&$-$0.700& 0.23&0.23&0.23 & 2 \\
    3273.067  &0.164& $-$0.300 &  0.63:&0.63&0.63   & 7 \\
    3279.266  &0.095& $-$0.230 & 0.43&0.13&0.33 & 2 \\
    3284.703 &0.000& $-$0.480 & 0.43:&0.38&0.33  & 2 \\
    3305.153 &0.039& $-$0.690 & 0.43&0.33&0.48  & 2 \\

\hline
\multicolumn{7}{c}{Continue} \\
\hline

\end{tabular}
\end{table}


\begin{table}
\renewcommand\thetable{A.1}  
\centering   
\scalefont{0.95}
\begin{tabular}{lccccccc}
     \midrule
    3314.488 &0.713&$-$0.686 & 0.33&0.33&0.33   & 2 \\
    3334.607 &0.559& $-$0.797& 0.43&0.43&0.33 & 2,6 \\
    3338.414&0.959&$-$0.578& 0.33 & 0.23 & 0.28  & 2,6 \\
    3340.574&0.164&$-$0.461& 0.43: & 0.00 & 0.03     & 2 \\
    3344.786 &1.011& $-$0.220 & 0.33 & 0.03 & 0.03 & 17 \\ 
    3356.088  &0.095& $-$0.513 & 0.43: & 0.33 & 0.43 & 2\\
    3357.264  &0.000& $-$0.736 & 0.43: & 0.33 & 0.43 & 2 \\
    3391.982&0.164&0.463& 0.42: & 0.23 & 0.33  & 2 \\
    3393.122&0.039&$-$0.700& 0.43 & 0.23 & 0.33 & 2 \\
    3402.868&1.532&$-$0.330& 0.23 & 0.23: & 0.23 & 2 \\
    3403.673&0.999&$-$0.576& 0.43 & 0.43 & 0.33 & 2 \\
    3408.096 &0.972& $-$0.596& 0.43 & 0.03 & 0.23 & 17 \\ 
     3410.236 &0.409& $-$0.323 & 0.23 & 0.23 & 0.03 & 6 \\
     3419.128 &0.164& $-$1.574 & 0.43 & 0.33 & 0.38 & 2  \\
     3424.813 &0.039& $-$1.305 &  0.43 & 0.33 & 0.38 & 2,5,6  \\
     3430.514  &0.466& $-$0.164 &  0.23 & 0.23 & 0.18 & 2,5,6 \\
     3438.226  &0.095& 0.310  & 0.43: & 0.03 & 0.18 & 6 \\
     3457.548  &0.559& $-$0.530 & 0.43 & 0.43 & 0.43 & 2,5,6 \\
     3458.920  &0.959& $-$0.520 & 0.33 & 0.23 & 0.23 & 6 \\
     3479.029  &0.527& $-$0.690 &  0.33 & 0.33 & 0.33 & 2,5,6 \\
     3479.383  &0.713& 0.120  & 0.33 & 0.23 & 0.33 & 2,5,6 \\
     3499.560 &0.409& $-$0.810 &  0.23 & 0.23 & 0.18  & 2,5,6 \\
     3505.682  &0.164& $-$0.360 & 0.43: & 0.33 & 0.23 & 2,5,6 \\
     3506.048  &1.236& $-$0.860 & 0.43 & 0.23 & 0.33 & 2 \\
     3520.869  &0.559& $-$1.089 & 0.23 & 0.03 & 0.23 & 2 \\
     3525.803  &0.359& $-$0.653 &  0.23 & 0.08 & 0.13 & 2  \\
     3536.935 &0.359& $-$1.306  & 0.33 & 0.33 & 0.28 & 2,5,6 \\
     3549.511 &1.236  & $-$0.400  & 0.33 & 0.23 & 0.23 & 17 \\  
     3551.939 &0.095& $-$0.310 &  0.33 & 0.13 & 0.23 & 2,6 \\
     3556.585  &0.466& 0.140  & 0.23 & 0.03 & 0.03& 2 \\
     3573.055 &0.319& $-$1.041 & 0.43: & 0.33 & 0.33  & 2,5,6 \\
     3578.205 &3.033& $-$1.596 & 0.43: & 0.33 & 0.33 & 2,5,6 \\
     3588.308  &0.409& $-$1.130 & 0.43 & 0.33 & 0.33 & 2 \\
     3607.373  &1.236& $-$0.640 & 0.33 & 0.23 & 0.28 & 2 \\
     3611.889&1.743&0.450& 0.23 & 0.23 & 0.03      & 2 \\
     3613.102&0.039&$-$0.465& 0.43: & -- & 0.23     & 2 \\
     3614.765  &0.359& $-$0.252 & 0.23 & 0.03 & 0.13 & 2 \\
     3630.004 &0.359& $-$1.110 & 0.33 & 0.33 & 0.33 & 2,5,6 \\
     3636.436&0.466&$-$1.035&  0.43 & 0.38 & 0.23      & 2 \\
     3674.696&0.319&$-$0.446& 0.33 & 0.23 & 0.13      & 2 \\
     3698.152&1.011&0.094       &0.23 & 0.18 & 0.13 & 6 \\
     3714.794 &0.527& $-$0.930    &0.63 & 0.53 & 0.53 & 2,5,6\\
     3751.606  &0.972& 0.012    &0.23 & 0.23 & 0.13 & 5,6 \\
     3766.795 &0.409& $-$0.812    &0.43 & 0.43 & 0.43 & 2,6\\
     3836.762  &0.559& $-$0.060   &0.23 & 0.03 & 0.03 & 1\\
     3998.954  &0.559& $-$0.387   &0.23: & 0.03 & 0.03 & 5,6 \\
    \midrule
    \multicolumn{7}{c}{{\ion{Nb}{II} (Z\,$=$\,41)}} \\
    \multicolumn{7}{c}{{log$\epsilon$(Nb) = A(Nb) = $-$0.65 }} \\
    \midrule
    3028.433  &0.439& $-$0.410 & $-$0.55: & -- & -- & 2  \\
    3191.093 &0.514& $-$0.260  & $-$0.55 & -0.75 & -- & 2  \\
    3215.591  &0.439& $-$0.190 & $-$0.55 & -0.65 & -0.75 & 1,5,6  \\
    3225.475 &0.292& $-$0.030  & $-$0.55: & -0.75:: & -0.65:: & 16  \\
    \midrule
    \multicolumn{7}{c}{{\ion{Mo}{I} (Z\,$=$\,42)}} \\
    \multicolumn{7}{c}{{log$\epsilon$(Mo) = A(Mo) = $-$0.21 }} \\
    \midrule
     3864.103& 0.000& $-$0.010  &$-$0.11 & -0.21 & -0.24 & 6  \\
    \midrule
    \multicolumn{7}{c}{{\ion{Ru}{I} (Z\,$=$\,44)}} \\
    \multicolumn{7}{c}{{log$\epsilon$(Ru) = A(Ru) = 0.18 }} \\
    \midrule
     3436.736  &0.148& 0.150  & 0.25 & 0.15 & 0.15 & 1,6  \\
     3498.942  &0.000& 0.310  & 0.25 & 0.15 & 0.15 & 1,6  \\
     3728.025  &0.000& 0.270  & 0.25 & 0.20 & 0.20 & 1,6  \\
     3798.898  &0.148& $-$0.040 & 0.25: & 0.20 & 0.20 & 1,6 \\
     3799.349  &0.000& 0.020  & 0.36: & 0.20 & 0.20 & 1,6 \\
    \midrule
    \multicolumn{7}{c}{{\ion{Rh}{I} (Z\,$=$\,45)}} \\
    \multicolumn{7}{c}{{log$\epsilon$(Rh) = A(Rh) = $-$0.51 }} \\
    \midrule
      3396.819  &0.000 & 0.050  & $-$0.42 & $-$0.62 & $-$0.42 & 1 \\
      3434.885  &0.000&  0.450  & $-$0.42 & $-$0.57 & $-$0.52  & 1,6 \\
      3692.358  &0.000&  0.174  & $-$0.42 & $-$0.42 & $-$0.52 & 1,6\\
      3700.907  &0.190 & $-$0.100 & $-$0.42 & $-$0.42: & $-$0.52: & 2 \\

\hline
\multicolumn{7}{c}{Continue} \\
\hline

\end{tabular}
\end{table}

\begin{table}
\renewcommand\thetable{A.1} 
\centering   
\scalefont{0.95}
\begin{tabular}{lccccccc}
    \midrule    
    \multicolumn{7}{c}{{\ion{Pd}{I} (Z\,$=$\,46)}} \\
    \multicolumn{7}{c}{{log$\epsilon$(Pd) = A(Pd) = $-$0.21 }} \\
    \midrule    
    3242.700   &0.814& $-$0.070 & $-$0.05 & $-$0.15 & $-$0.15 & 1,5,6 \\
    3404.579  &0.814&  0.320 &  $-$0.20 & $-$0.35 & $-$0.35  & 1,5,6 \\
    3460.739  & 0.814 & $-$0.420& $-$0.15 & $-$0.20 & $-$0.20  & 6 \\
    3516.944  &0.962& $-$0.240 &  $-$0.05 & $-$0.20 & $-$0.10 & 2,5,6 \\
    3634.690  &0.814&  0.090 & $-$0.05 & $-$0.15 & $-$0.05  & 1 \\
    \midrule
    \midrule
    \multicolumn{7}{c}{{\ion{Ag}{I} (Z\,$=$\,47)}} \\
    \multicolumn{7}{c}{{log$\epsilon$(Ag) = A(Ag) = $-$0.94 }} \\
    \midrule    
    3280.679  &0.000& $-$0.050 & $-$1.01 & $-$1.01 & $-$0.96  & 1,5,6 \\
    3382.889   &0.000& $-$0.377 & $-$0.91 & $-$0.87 & $-$0.91  & 1,5,6 \\
    \midrule
    \multicolumn{7}{c}{{\ion{Cd}{I} (Z\,$=$\,48)}} \\
    \multicolumn{7}{c}{{log$\epsilon$(Cd) = A(Cd) = --- }} \\
    \midrule    
    3261.050  &0.000& -2.470  & ---  & --- & --- & 6 \\
    \midrule
    \multicolumn{7}{c}{{\ion{Sn}{I} (Z\,$=$\,50)}} \\
    \multicolumn{7}{c}{{log$\epsilon$(Sn) = A(Sn) = $<-$0.40 }} \\
    \midrule     
    3262.331  &1.068& 0.110 &  $-$0.90 & <-0.40 & -- & 6 \\
    3655.790  &2.128& $-$0.450 &  --- & ... \\ 
    3801.011  &1.068& $-$0.620 &  --- & 6 \\
    \midrule
    \multicolumn{7}{c}{{\ion{Ba}{II} (Z\,$=$\,56)}} \\
    \multicolumn{7}{c}{{log$\epsilon$(Ba) = A(Ba) = 0.40 - see text }} \\
    \midrule 
    3891.776$^{*}$  &2.512&  0.280 & +0.15 & +0.15 & 0.00 & 1,5,6 \\
    \midrule
    \multicolumn{7}{c}{{\ion{La}{II} (Z\,$=$\,57)}} \\ 
    \multicolumn{7}{c}{{log$\epsilon$(La) = A(La) = $-$0.73 }} \\
    \midrule  
    3713.545$^{*}$ &0.173& $-$0.800 & $-$0.60 & $-$0.60 & $-$0.60 & 5,6,9 \\ 
    3794.774$^{*}$ &0.244& $-$0.443 & $-$0.60 & $-$0.80 & $-$0.90 & 5,6,9 \\
    3849.006$^{*}$ &0.000& $-$0.450 & $-$0.60 & $-$0.75 & $-$0.80 & 1,5,9 \\
    3949.102$^{*}$ &0.403& $-$1.690 & $-$0.80 & $-$0.80 & $-$0.60 & 5,6,9 \\ 
    3988.515$^{*}$ &$-$1.015& 0.210 & $-$0.70 & $-$0.70 & $-$0.70 & 5,6,9 \\
    3995.745$^{*}$ &$-$1.109& 0.060 & $-$0.60 & $-$0.75 & $-$0.70 & 5,6,9 \\
    \midrule
    \multicolumn{7}{c}{{\ion{Ce}{II} (Z\,$=$\,58)}} \\
    \multicolumn{7}{c}{{log$\epsilon$(Ce) = A(Ce) = $-$0.41 }} \\
    \midrule 
    3263.885 & 0.459 & $-$0.390 & -- & -- & -- & 19,2\\ 
    3426.205 &0.122& $-$0.660 & $-$0.31 & $-$0.31 & $-$0.41  &  2,6,19\\
    3507.941 &0.175& $-$0.960 & $-$0.31 & $-$0.15 & $-$0.31 & 2,19 \\
    3520.520  &0.175& $-$0.910 &  $-$0.31 & $-$0.25 & $-$0.31 & 2,19 \\
    3534.045 &0.521& $-$0.140 &  $-$0.31 & $-$0.36 & $-$0.26 &  2,19 \\
    3539.079 &0.320& $-$0.270 & $-$0.31 & $-$0.41 & $-$0.41  &  2,5,6,19\\
    3577.456 &0.470&  0.140 &  $-$0.31 & $-$0.41 & $-$0.31 &  2,5,6,19 \\  
    3655.844 & 0.318 & $-$0.050  & $-$0.31 & $-$0.41 & $-$0.46  & 6,19\\
    3659.225 &0.175& $-$0.670 & $-$0.31 & $-$0.41 & $-$0.36  & 2,19 \\
    3709.929 &0.122& $-$0.260 &  $-$0.31 & $-$0.41 & $-$0.41 & 2,19 \\
    3781.616 &0.529& $-$0.260 &  $-$0.31 & $-$0.41 & $-$0.36 & 2,19 \\
    3940.330 & 0.318&$-$0.270 & $-$0.31 & $-$0.41 & $-$0.36  & 5,6,19 \\
    3940.660 & 0.495&$-$0.991 &  $-$0.31 & -- & --  & 19 \\
    3940.970 & 0.417&$-$0.570 &  $-$0.31 & -- & -- & 19 \\
    3942.151 &0.000& $-$0.220 & $-$0.31 & $-$0.41 & $-$0.41  & 5,6,19 \\ 
    3942.745 &0.857&  0.690 & $-$0.31 & $-$0.61 & $-$0.61  & 5,6,19 \\ 
    3960.909 &0.322& $-$0.360 &  $-$0.31 & $-$0.41 & $-$0.41 & 5,6,19 \\ 
    3964.496 &0.322&$-$0.650  &  $-$0.31 & $-$0.51 & $-$0.61 & 6,19\\
    3984.671 &0.956& 0.000  &-- & -- & -- & 6,19\\
    3992.380 &0.446& $-$0.220 & $-$0.31 & $-$0.41 & $-$0.41  & 6,19\\
    3999.237  &0.090& 0.060 & -- & $-$0.61 & -- & 5,6 \\
    \midrule
    \multicolumn{7}{c}{{\ion{Pr}{II} (Z\,$=$\,59)}} \\
    \multicolumn{7}{c}{{log$\epsilon$(Pr) = A(Pr) = $-$0.94 }} \\
    \midrule 
    3964.262$^{*}$ &0.215& $-$0.230 & $-$0.86 & -- & -- & 1 \\
    3964.812$^{*}$ &0.055& $-$0.920 & $-$0.86 & $-$0.96 & $-$0.96 & 1,5,6 \\ 
    3965.253$^{*}$ &0.204& $-$0.854 & $-$0.86 & $-$0.91 & $-$0.91 & 1,5,6 \\ 
    \midrule    

\hline
\multicolumn{5}{c}{Continue} \\
\hline

\end{tabular}
\end{table}


\begin{table}
\renewcommand\thetable{A.1}  
\centering   
\scalefont{0.95}
\begin{tabular}{lccccccc}
    \midrule
    \multicolumn{7}{c}{{\ion{Nd}{II} (Z\,$=$\,60)}} \\
    \multicolumn{7}{c}{{log$\epsilon$(Nd) = A(Nd) = $-$0.33 }} \\
    \midrule 
    3285.085 & 0.000 & $-$1.050 & -- & -- & --& 2 \\
    3300.143 & 0.000 & $-$1.036 & $-$0.13 & $-$0.33 & $-$0.33 & 2 \\
    3325.889 &0.064& $-$1.174 & $-$0.13 & $-$0.23 & $-$0.13  & 2  \\
    3334.465 &0.182& $-$0.930 & -- & -- & -- & 2 \\
    3555.764 &0.321& $-$0.950 & $-$0.13 & $-$0.23 & $-$0.23 & 2 \\
    3560.718 &0.471& $-$0.500 & $-$0.13 & $-$0.33 & $-$0.43 & 2 \\
    3598.021 &0.064& $-$1.020 & $-$0.13 & $-$0.23 & $-$0.18 & 2 \\
    3609.780 &0.000& $-$0.800 & $-$0.13 & $-$0.28 & $-$0.33 & 2 \\
    3730.577 &0.380& $-$0.611 & $-$0.13 & $-$0.20 & $-$0.23 & 2 \\
    3738.055 & 0.559 & $-$0.040 & $-$0.13 & $-$0.23 & $-$0.23 & 2 \\
    3741.424 & 0.064 & $-$0.680 & $-$0.13 & $-$0.33 & $-$0.33 & 2 \\
    3763.472 & 0.205 & $-$0.430 & $-$0.13 & $-$0.23 & $-$0.33 & 2 \\
    3779.462 & 0.182 & $-$0.560 & $-$0.13 & $-$0.33 & $-$0.33 & 2 \\
    3780.382 &0.471& $-$0.350 & $-$0.13 & $-$0.23 & $-$0.33 & 2,5,6 \\ 
    3784.245 &0.380&  0.150 & $-$0.13 & $-$0.33 & $-$0.33 & 2,5,6 \\
    3795.454 & 0.205 & $-$0.650 & $-$0.13 & $-$0.28 & $-$0.28 & 2 \\
    3803.471 & 0.205 & $-$0.390 & $-$0.13 & $-$0.33 & $-$0.33 & 2 \\
    3808.767 & 0.064 & $-$0.650 & $-$0.13 & $-$0.53 & $-$0.53 & 2 \\
    3810.477   &0.742& $-$0.140 & $-$0.13 & $-$0.38 & $-$0.33 & 14 \\
    3826.409   &0.064& $-$0.410 & $-$0.13 & $-$0.33 & $-$0.33 & 5,6 \\
    3838.981  &0.000& $-$0.240 & $-$0.13 & $-$0.43 & $-$0.48 & 5,6 \\  
    3865.956  &0.380& $-$1.048 & $-$0.13 & $-$0.43 & $-$0.43 & 5,6 \\
    3866.791 & 0.205 & $-$0.950 & $-$0.13 & $-$0.53 & $-$0.53 & 6 \\
    3900.219  &0.471&  0.100 & $-$0.13 & $-$0.33 & $-$0.33 & 5,6 \\
    3973.260& 0.631 & 0.360 &  $-$0.13 & $-$0.43 & $-$0.43 & 1,5,6 \\ 
    \midrule
    \multicolumn{7}{c}{{\ion{Sm}{II} (Z\,$=$\,62)}} \\
    \multicolumn{7}{c}{{log$\epsilon$(Sm) = A(Sm) = $-$0.54 }} \\
    \midrule     
    3218.596 &0.185& $-$0.793 & $-$0.51 & $-$0.31 & $-$0.51 & 2 \\
    3244.686 &0.185& $-$1.399 & $-$0.11: & $-$0.21: & -- & 2 \\
    3253.403 &0.104& $-$1.080 & $-$0.31 & $-$0.31 & $-$0.31 & 2 \\
    3304.517 &0.000& $-$1.190 & $-$0.51 & $-$0.51 & $-$0.51: & 2 \\
    3307.027 &0.659& $-$0.301 & $-$0.11 & $-$0.21 & $-$0.11 & 2 \\
    3321.189 &0.378& $-$0.362 & $-$0.51 & $-$0.51 & $-$0.51 & 2 \\
    3384.654 &0.378& $-$0.741 & $-$0.51 & $-$0.61 & $-$0.46 & 2 \\
    3568.271 &0.485&  0.298 & $-$0.51 & $-$0.51 & $-$0.71 & 2 \\
    3583.372 &0.185& $-$1.119 & $-$0.51 & $-$0.51 & $-$0.31 & 2 \\
    3604.281 &0.485& $-$0.158 & $-$0.51 & $-$0.51 & $-$0.41 & 2 \\
    3609.492 & 0.277 & 0.160 & $-$0.71 & $-$0.81 & $-$0.71 & 2 \\
    3621.210 &0.104& $-$0.442 & $-$0.51 & $-$0.91 & $-$0.71 & 2 \\
    3627.004 &0.277& $-$0.614 & $-$0.51 & $-$0.51 & $-$0.51 & 2 \\
    3661.352 &0.041& $-$0.427 & $-$0.51 & $-$0.91 & $-$0.71 & 2 \\
    3670.821 &0.104& $-$0.344 & $-$0.51 & $-$0.71 & $-$0.71 & 2 \\
    3706.752 & 0.485 & $-$0.600 & $-$0.51 & $-$0.51 & $-$0.51 & 2,6 \\
    3718.883 & 0.378 & $-$0.310 & $-$0.51 & $-$0.61 & $-$0.91 & 2 \\
    3731.263 &0.104& $-$0.384 & $-$0.51 & $-$0.71 & $-$0.61 & 2 \\
    3739.120 &0.041& $-$0.846 & $-$0.51 & $-$0.41 & $-$0.51:: & 2 \\
    3743.877 &0.333& $-$0.428 & $-$0.51 & $-$0.51 & $-$0.41 & 2 \\
    3758.460 &0.000& $-$1.102 & $-$0.51 & $-$0.31 & $-$0.71 & 2 \\
    3760.710 &0.185& $-$0.428 & $-$0.51 & $-$0.61 & $-$0.51 & 2,6 \\
    3762.588 &0.248& $-$0.751 & $-$0.51 & $-$0.61 & $-$0.61 & 2 \\
    3793.978& 0.104 & $-$0.498 & $-$0.63 & $-$0.58 & $-$0.71 & 1 \\ 
    3896.972$^a$  &0.041& $-$0.578 & $-$0.51 & $-$0.51 & $-$0.71 & 1,5,6 \\
    3993.309  &0.041  & $-$0.894  & $-$0.51 & $-$0.51 & $-$0.61 &  6\\
    \midrule
    \multicolumn{7}{c}{{\ion{Eu}{II} (Z\,$=$\,63)}} \\
    \multicolumn{7}{c}{{log$\epsilon$(Eu) = A(Eu) = $-$0.93}} \\
    \midrule 
    3688.430$^{*}$  &0.000&  $-$0.670 & $-$0.76 & $-$0.96 & $-$1.16 & 10, 20 \\ 
    3724.930$^{*}$ &0.000& $-$0.090 & $-$0.76   & $-$0.76 & $-$0.86 & 1,5,6,10,20 \\ 
    3819.672$^{*}$  &0.000& 0.510  & $-$0.76 & $-$0.96 & $-$0.96 & 5,6,10,20 \\ 
    3907.107$^{*}$  &0.207& 0.170  & $-$0.76 & $-$0.96 & $-$0.96 & 5,6,10,20 \\ 
    3930.499$^{*}$ &0.207& 0.270  & $-$0.76 & $-$0.96 & $-$0.86 & 1,5,6,10,20 \\ 
    3971.972$^{*}$  &0.207&  0.270 & $-$0.86 & $-$0.96 & $-$0.96 & 1,5,6,10,20 \\ 

 \hline
\multicolumn{5}{c}{Continue} \\
\hline

\end{tabular}
\end{table}


\begin{table}
\renewcommand\thetable{A.1}
\centering  
\scalefont{0.97}
\begin{tabular}{lccccccc}
    \midrule
    \multicolumn{7}{c}{{\ion{Gd}{II} (Z\,$=$\,64)}} \\
    \multicolumn{7}{c}{{log$\epsilon$(Gd) = A(Gd) = $-$0.45 }} \\
    \midrule 

    3032.844 & 0.079 &  0.300& $-$0.47 & -- & -- & 8 \\ 
    3034.051 & 0.032 &  0.149& $-$0.47: & -- & --  &   8  \\ 
    3076.928 & 0.000& $-$0.480& $-$0.27 & $-$0.27 & --  &   8  \\ 
    3100.504 & 0.240&  0.620& $-$0.27 & $-$0.42 & --  &   8  \\ 
    3124.262 & 0.032& $-$1.250& $-$0.27 & $-$0.47 & --  &   8  \\ 
    3331.387 & 0.000& $-$0.140  & $-$0.27 & $-$0.47 & $-$0.57 & 5,6 \\
    3358.625 &0.032& 0.152   & $-$0.27 & $-$0.27 & $-$0.27  & 2 \\
    3360.712 &0.032& $-$0.240  & -- & -- & -- & 2 \\
    3362.239 &0.079& 0.294   & $-$0.27 & $-$0.37 & $-$0.57: & 2 \\
    3364.245 &0.000& $-$1.086  & $-$0.27 & $-$0.57 & $-$0.67: & 2 \\
    3392.527 &0.079& $-$0.220  & $-$0.27 & $-$0.47 & $-$0.47 & 2 \\
    3418.729 &0.000& $-$0.310  & $-$0.27 & $-$0.47 & $-$0.47 & 2 \\
    3422.464 &0.240& 0.519   & +0.03 & $-$0.77 & $-$0.27 & 2 \\
    3423.924 &0.000& $-$0.520  & $-$0.27 & $-$0.37 & $-$0.47 & 2 \\
    3424.595 &0.354&$-$0.170   & $-$0.27 & $-$0.47 & $-$0.47 & 6 \\
    3439.208 &0.382& 0.150   & $-$0.42 & $-$0.47 & $-$0.47 & 2,5,6 \\
    3439.787 &0.425& $-$0.230  & $-$0.12 & $-$0.27 & $-$0.17 & 2 \\
    3439.988 &0.240& 0.100   & $-$0.12 & $-$0.27 & $-$0.17 & 2 \\
    3451.236 &0.382& $-$0.050  & $-$0.47 & $-$0.57 & $-$0.47 & 2,5,6 \\
    3454.907 &0.032& $-$0.480  & $-$0.27 & $-$0.47 & $-$0.52 & 2,6 \\
    3463.990  &0.427& 0.269  & $-$0.27 & $-$0.47 & $-$0.27 & 2 \\
    3467.274 &0.425& 0.064  & $-$0.27 & $-$0.47 & $-$0.52 & 2 \\
    3473.224 &0.032& $-$0.412 & $-$0.27 & $-$0.27 & $-$0.27 & 2 \\
    3481.802 &0.492& 0.230  & $-$0.27 & $-$0.47 & $-$0.47 & 2 \\
    3482.607 &0.427& $-$0.484 & -- & -- & --  & 2 \\
    3491.960  &0.000& $-$0.611 & $-$0.27 & $-$0.27 & $-$0.22 & 2 \\
    3549.359   &0.240& 0.260 & $-$0.27 & $-$0.47 & $-$0.47 & 5,6 \\
    3557.058  &0.600& 0.210 & $-$0.27 & $-$0.47 & $-$0.47 & 2,5,6 \\
    3646.196 &0.240& 0.328  & $-$0.27 & $-$0.47 & $-$0.47 & 2 \\
    3654.624 &0.079& $-$0.030 & $-$0.27 & $-$0.47 & $-$0.47 & 2 \\
    3656.152 &0.144& $-$0.067 & $-$0.27 & $-$0.42 & $-$0.47 & 2 \\
    3671.205 &0.079& $-$0.330 & $-$0.27 & $-$0.47: & $-$0.67 & 2 \\
    3697.733   &0.032& $-$0.280 & $-$0.27 & $-$0.37 & $-$0.27 & 5,6 \\
    3699.737 &0.354& $-$0.260 & $-$0.27 & $-$0.47 & $-$0.47 & 2 \\
    3712.704   &0.382& 0.150 & $-$0.47 & $-$0.57 & $-$0.57 & 5,6 \\
    3768.396 &0.079& 0.360  & $-$0.27 & $-$0.57 & $-$0.57 & 1,5,6\\
    3796.384 &0.032&  0.14  & $-$0.27 & $-$0.47 & $-$0.47 & 1,6\\
    3836.915&0.492 & $-$0.322 & $-$0.27 & $-$0.47 & $-$0.37 & 1 \\ 
    3844.578 &0.144& $-$0.400 & $-$0.27 & $-$0.37 & $-$0.37 & 1,5,6\\
    3916.509& 0.600 & 0.060 &  $-$0.27 & $-$0.47 & $-$0.47 & 1 \\ 
    3973.977 &0.602 & $-$0.400 & $-$0.47 & $-$0.57 & $-$0.67 & 6 \\
    \midrule
    \multicolumn{7}{c}{{\ion{Tb}{II} (Z\,$=$\,65)}} \\ 
    \multicolumn{7}{c}{{log$\epsilon$(Tb) = A(Tb) = $-$1.22 }} \\
    \midrule 
    3070.060 & 0.400 & 0.170 &  $-$1.26 & $-$1.00 & -- & 8 \\ 
    3472.800$^{*}$  & 0.126 & $-$0.10 & $-$1.26: & $-$1.26: & $-$1.36: & 6,11,12 \\
    3509.144 &0.000& 0.700  &   $-$1.26 & $-$1.26 & $-$1.26 & 2,11,12 \\
    3568.510$^{*}$   &0.000& $-$1.928 &  $-$1.26 & $-$1.26 & $-$1.26 & 1,6,11,12 \\ 
    3600.410$^{*}$ &0.641& $-$1.519  &  $-$1.26 & $-$1.26 & $-$1.16 & 5,6,11,12 \\
    3633.287 & 0.641 & 0.090  &  $-$1.26: & $-$1.26: & $-$1.26:  & 2,11,12 \\
    3641.655 &0.649& 0.040  &   $-$1.26: & $-$1.26: & $-$1.26: & 2,11,12 \\
    3658.886$^{*}$ &0.126& -4.083  &   $-$1.26 & $-$1.26 & $-$1.26 & 1,5,6,11,12\\
    3702.853$^{*}$  &0.126& $-$1.794 &   $-$1.26 & $-$1.26 & $-$1.16 & 1,5,6,11,12 \\
    3848.734$^{*}$ &0.000& -2.008  &   $-$1.26 & $-$1.26 & $-$1.26 & 1,5,6,11,12 \\
    3874.168$^{*}$ &0.000& $-$0.317 &   -- & $-$1.15 & $-$1.26  & 1,5,6,11,12 \\ 
    3899.188& 0.373 & 0.330 &   $-$1.26 & $-$1.15 & $-$1.00   & 1,11,12 \\

\hline
\multicolumn{5}{c}{Continue} \\
\hline

\end{tabular}
\end{table}


\begin{table}
\renewcommand\thetable{A.1}
\centering  
\scalefont{0.97}
\begin{tabular}{lccccccc}
    \midrule
    \multicolumn{7}{c}{{\ion{Dy}{II} (Z\,$=$\,66)}} \\
    \multicolumn{7}{c}{{log$\epsilon$(Dy) = A(Dy) = $-$0.25 }} \\
    \midrule  
    3026.160 & 0.000 & $-$0.980& $-$0.50 & -- & -- & 8 \\ 
    3407.796 &0.000& 0.180   & $-$0.21 & $-$0.21 & $-$0.21 & 2,5,6 \\
    3413.784 &0.103& $-$0.460  & $-$0.21 & $-$0.21 & $-$0.21 & 2 \\
    3434.369 &0.000& $-$0.450  & $-$0.21 & $-$0.21 & $-$0.21 & 2,5,6 \\
    3445.574  &0.000& $-$0.150 & $-$0.21 & $-$0.21 & $-$0.21 & 5,6 \\
    3449.892& 0.538& $-$0.524 & $-$0.21 & $-$0.41 & $-$0.26  & 6 \\
    3454.317 &0.103& $-$0.140  & $-$0.21 & $-$0.21 & $-$0.41 & 2,5,6 \\
    3456.559 &0.590& $-$0.007  & $-$0.21 & $-$0.41 & $-$0.41 & 2 \\
    3460.969 &0.000& $-$0.070  & $-$0.21 & $-$0.21 & $-$0.31 & 2,5,6 \\
    3473.697 &0.928 & $-$0.218 & $-$0.21 & $-$0.41 & $-$0.41 & 6 \\
    3506.815 &0.103 & $-$0.440  & $-$0.21 & $-$0.36 & $-$0.36 & 6 \\
    3531.707  &0.000& 0.790   & $-$0.21 & $-$0.21 & $-$0.41 & 2,6 \\
    3534.960 & 0.103 & $-$0.040 & $-$0.21 & $-$0.21 & $-$0.21  & 2 \\
    3536.019 & 0.538 & 0.530 &  $-$0.21 & $-$0.21 & $-$0.41   & 2,5,6 \\
    3538.519 &0.000& $-$0.020  & $-$0.21 & $-$0.11 & $-$0.21 & 5,6 \\
    3546.832 &0.103& $-$0.550  &  $-$0.21 & $-$0.21 & $-$0.21  & 2,5,6 \\
    3550.218  &0.590& 0.270  & $-$0.21 & $-$0.36 & $-$0.31 & 2,5,6 \\
    3559.295  &1.224& $-$0.280 & $-$0.21 & $-$0.21 & $-$0.41 & 5,6 \\
    3563.148  &0.103& $-$0.360  & $-$0.21 & $-$0.21 & $-$0.21 & 2,5,6 \\
    3640.249 &0.590 & $-$0.370 & $-$0.21 & $-$0.21 & $-$0.31 & 6 \\
    3694.810  &0.103& $-$0.110  & $-$0.21 & $-$0.21 & $-$0.41 & 2,5,6 \\
    3708.221 &0.590 & $-$0.880 & $-$0.21 & $-$0.21 & $-$0.21 & 6 \\
    3747.817 &0.103& $-$0.810 & $-$0.21 & $-$0.21 & $-$0.21 & 5,6 \\
    3788.436 &0.103& $-$0.570 & $-$0.21 & $-$0.21 & $-$0.21 & 5,6 \\
    3869.864& 0.000 & $-$1.050 & $-$0.21 & $-$0.21 & $-$0.26 & 1 \\ 
    3944.681  &0.000& 0.100  & $-$0.21 & $-$0.36 & $-$0.21 & 7 \\
    3983.651 &0.538& $-$0.310 & $-$0.21 & $-$0.21 & $-$0.21  & 5,6 \\
    3996.689  &0.590& $-$0.260 & $-$0.21 & $-$0.21 & $-$0.21 & 1,5,6 \\
    \midrule
    \multicolumn{7}{c}{{\ion{Ho}{II} (Z\,$=$\,67)}} \\ 
    \multicolumn{7}{c}{{log$\epsilon$(Ho) = A(Ho) = $-$0.98 }} \\
    \midrule 
    3796.748$^{*}$ &0.000& 0.160 & $-$1.0 & $-$0.96 & $-$0.96 & 13  \\ 
    3810.738$^{*}$ &0.000& 0.190 & $-$0.9 & $-$0.91 & $-$0.91 & 5,6,13\\ 
    3890.925$^{*}$ &$-$0.293& 0.460 & $-$1.0 & $-$1.06 & $-$1.06 & 5,13\\ 
    3905.634 & 0.079& $-$0.530 & -- & -- & -- & 6,13 \\ 
    \midrule
    \multicolumn{7}{c}{{\ion{Er}{II} (Z\,$=$\,68)}} \\
    \multicolumn{7}{c}{{log$\epsilon$(Er) = A(Er) = $-$0.38 }} \\
    \midrule 
    3028.275& 0.00 &  1.02 & -- & -- & -- & 8 \\ 
    3073.344 & 0.000 & $-$0.610 & $-$0.42 & $-$0.47 & -- & 8 \\ 
    3332.703 & 0.886 & 0.041 & $-$0.27 & $-$0.47 & --  & 6 \\
    3364.076 &0.055& $-$0.487  & $-$0.32 & $-$0.42 & $-$0.27   & 2  \\
    3441.130  &0.055& $-$0.672 & $-$0.27 & $-$0.27 & $-$0.27  & 2  \\
    3499.103 &0.055& 0.139   & $-$0.27 & $-$0.47 & $-$0.47   & 2,5  \\
    3524.913 &0.000& $-$0.887  & $-$0.27 & $-$0.37 & $-$0.37   & 2 \\
    3549.844 &0.670& $-$0.310  & $-$0.27 & $-$0.47 & $-$0.47  & 2 \\
    3559.894 &0.000& $-$0.736  & $-$0.32 & $-$0.37 & $-$0.37  & 2,5,6 \\
    3580.518 &0.055& $-$0.768  & $-$0.27 & $-$0.27 & $-$0.27  & 2 \\
    3616.566 &0.000& $-$0.327  & $-$0.27 & $-$0.27 & $-$0.27  & 2 \\
    3618.916 &0.670& $-$0.594  & $-$0.27 & $-$0.07: & $-$0.27:  & 2 \\
    3633.536 &0.000& $-$0.694  & $-$0.27 & $-$0.37 & $-$0.42  & 2 \\
    3692.649  &0.055& 0.138  & $-$0.27 & $-$0.37 & $-$0.37  & 1,2,5,6 \\
    3700.720  &0.055& $-$1.290 & $-$0.27 & $-$0.27 & $-$0.27  & 2 \\
    3729.524  &0.000& $-$0.488 & $-$0.37 & $-$0.47 & $-$0.47  & 2,6 \\
    3742.640  &0.636& $-$0.474 & $-$0.27 & $-$0.47 & $-$0.47  & 2 \\
    3786.836  &0.000& $-$0.644 & $-$0.27 & $-$0.27 & $-$0.47  & 1,2,5 \\
    3830.482  &0.000& $-$0.365 & $-$0.27 & $-$0.42 & $-$0.47 & 1,5 \\
    3896.234  &0.055& $-$0.241 & $-$0.27 & $-$0.42 & $-$0.42 & 1,5 \\
    3906.312  &0.000& $-$0.052 & $-$0.27 & $-$0.47 & $-$0.47 & 7 \\
    3938.626 &0.000& $-$0.520  & $-$0.27 & $-$0.47 & $-$0.47 & 1,5,6 \\

\hline
\multicolumn{7}{c}{Continue} \\
\hline

\end{tabular}
\end{table}


\begin{table}
\renewcommand\thetable{A.1}
\centering  
\scalefont{0.97}
\begin{tabular}{lccccccc}
    \midrule
    \multicolumn{7}{c}{{\ion{Tm}{II} (Z\,$=$\,69)}} \\
    \multicolumn{7}{c}{{log$\epsilon$(Tm) = A(Tm) = $-$1.32 }} \\
    \midrule 
    3015.294 & 0.029 & $-$0.590 & $-$1.24 & -- & -- & 2 \\
    3131.255 &0.000& 0.240  & $-$1.24 & $-$1.44 & -- & 2 \\
    3362.615 &0.029& $-$0.100 & $-$0.89: & $-$1.04 & $-$1.04 & 2 \\
    3397.498 &0.000& $-$0.750 & $-$1.24 & $-$1.24 & $-$1.14 & 2 \\
    3462.197 &0.000& 0.030  & $-$1.24 & $-$1.39 & $-$1.34 & 2,6 \\ 
    3700.256 &0.029& $-$0.290 & $-$1.24 & $-$1.24 & $-$1.24  & 1,2,5,6 \\ 
    3701.363  &0.000& $-$0.540 & $-$1.24 & $-$1.24 & $-$1.04 & 2,6 \\ 
    3761.914 &0.000& $-$0.430 & $-$1.39 & $-$1.64 & $-$1.54  & 1,2,6 \\ 
    3795.760  &0.029& $-$0.230 & $-$1.24 & $-$1.29 & $-$1.34 & 1,2,6 \\ 
    3848.020  &0.000& $-$0.130 & $-$1.24 & $-$1.39 & $-$1.34 & 1,5,6 \\ 
    \midrule
    \multicolumn{7}{c}{{\ion{Yb}{II} (Z\,$=$\,70)}} \\ 
    \multicolumn{7}{c}{{log$\epsilon$(Yb) = A(Yb) = $-$0.70 }} \\
    \midrule 
    3289.367$^{*}$   &0.000& $-$0.050  & $-$0.65  & $-$0.70 & $-$0.40 & 6,19 \\ 
    3694.192$^{*}$   &5.724&  $-$0.200 & $-$0.70  & $-$0.70 & $-$0.85 & 6,19 \\ 
    \midrule
    \multicolumn{7}{c}{{\ion{Lu}{II} (Z\,$=$\,71)}} \\ 
    \multicolumn{7}{c}{{log$\epsilon$(Lu) = A(Lu) = $-$1.14 }} \\
    \midrule 
    3077.605$^{*}$ &1.541& $-$0.653 &  $-$1.14 & $-$1.14 & -- & 2 \\ 
    3397.066$^{*}$ &1.462& $-$1.006 & $-$1.14 & $-$1.14 & $-$0.94 & 6 \\
    3472.477$^{*}$ &1.541& $-$1.412 & -- & -- & -- & 6 \\ 
    3554.416$^{*}$ &2.148& $-$0.898 & -- & -- & -- & 6 \\ 
 \midrule
    \multicolumn{7}{c}{{\ion{Hf}{II} (Z\,$=$\,72)}} \\
    \multicolumn{7}{c}{{log$\epsilon$(Hf) = A(Hf) = $-$0.88 }} \\
    \midrule     
    3012.900 & 0.000 & $-$0.600 & $-$0.59 & -- & -- & 2 \\
    3109.113 &0.787& $-$0.250 & $-$0.59 & $-$0.89 & -- & 2 \\ 
    3255.279 &0.452& $-$1.130 & $-$0.59 & $-$0.89 & $-$0.74 & 2 \\ 
    3399.793 &0.000& $-$0.570 & $-$0.89 & $-$0.95 & $-$0.89 & 1,2 \\ 
    3569.034 &0.787& $-$0.400 & $-$0.59 & $-$0.89 & $-$0.89  & 2 \\
    3719.276  &0.608& $-$0.810 & $-$0.59 & $-$0.89 & $-$0.89 & 1,6 \\ 
    3793.379 & 0.378 & $-$1.110 & -- & $-$0.79: & $-$0.84 & 6 \\
    \midrule
    \multicolumn{7}{c}{{\ion{Os}{I} (Z\,$=$\,76)}} \\
    \multicolumn{7}{c}{{log$\epsilon$(Os) = A(Os) = 0.23 }} \\
    \midrule  
    3018.036 &0.000& $-$0.720 & 0.10 & -- & --  & 5 \\
    3058.655 &0.000& $-$0.451 & 0.30 & 0.10 & -- & 2,3,5,6 \\ 
    3267.945 &0.000& $-$1.080 & 0.20 & 0.10 & 0.10   & 5,6  \\
    3301.565 &0.000& $-$0.743 & 0.20 & 0.20 & 0.15  & 5,6 \\
    3528.598 & 0.000 & $-$1.740 & -- & 0.50 & 0.30 &  6 \\
    \midrule
    \multicolumn{7}{c}{{\ion{Ir}{I} (Z\,$=$\,77)}} \\
    \multicolumn{7}{c}{{log$\epsilon$(Ir) = A(Ir) = 0.10 }} \\
    \midrule  
    3047.158 & 1.623 & $-$0.500 & 0.20: & -- & -- & 3 \\
    3220.776   &0.352& $-$0.510 & 0.50: & 0.30: & 0.40: & 5\\ 
    3513.648  &0.000& $-$1.260 & 0.20 & 0.10 & 0.20 & 1,5,6\\
    3558.993 & 0.717 & $-$1.670 & 0.00 & 0.00 & 0.00 & 6 \\
    3800.120  &0.000& $-$1.450 & 0.20 & 0.00 & 0.20 & 1,5,6\\
    \midrule
    \multicolumn{7}{c}{{\ion{Pt}{I} (Z\,$=$\,78)}} \\ 
    \multicolumn{7}{c}{{log$\epsilon$(Pt) = A(Pt) = 0.0 (see text)}} \\
    \midrule  
    3064.711$^{*}$ &0.000& $-$0.340 & $-$0.50 & $-$0.50 & -- & 5,21 \\
    3139.387 & 0.100 & $-$1.580 & 0.50 & 0.50 & --  &  8 \\ 
    3301.859$^{*}$ & 0.814 & $-$0.770  &$-$1.10 & $-$1.10 & $-$1.10 &  6,21 \\
    3315.042 & 0.000 & -2.580 & 0.50 & 0.50 & -- &  6 \\
    \midrule
    \multicolumn{7}{c}{{\ion{Pb}{I} (Z\,$=$\,82)}} \\
    \multicolumn{7}{c}{{log$\epsilon$(Pb) = A(Pb) = $-$0.65 (see text) }} \\
   \midrule  
    3639.568  &0.969& $-$0.720  & --- & --- & --- &5 \\
    3683.462  &0.969& $-$0.600 & --- & --- & --- & 5,6 \\

\hline
\multicolumn{5}{c}{Continue} \\
\hline
\end{tabular}
\end{table}


\begin{table}
\renewcommand\thetable{A.1}
\centering  
\scalefont{0.97}
\begin{tabular}{lccccccc}

    \midrule
    \multicolumn{7}{c}{{\ion{Bi}{I} (Z\,$=$\,83)}} \\
    \multicolumn{7}{c}{{log$\epsilon$(Bi) = A(Bi) = $-$0.20 (see text) }} \\
    \midrule
    3024.635  &1.914& -0.15 &  $-$0.20 & --- & --- & 3 \\
    3067.707 & 0.000 & 0.220 &  --- & --- & --- & 3 \\
    \midrule
    \multicolumn{7}{c}{{\ion{Th}{II} (Z\,$=$\,90)}} \\
    \multicolumn{7}{c}{{log$\epsilon$(Th) = A(Th) = $-$1.04 }} \\
    \midrule
    3180.194  &0.189& $-$0.547  & -- & -- & -- & 15 \\ 
    3351.229  &0.188& $-$0.600  & $-$0.98 & $-$0.98 & $-$0.98  & 1 \\
    3433.999  &0.230& $-$0.537  & $-$0.98 & $-$1.18 & $-$1.08  & 1 \\ 
    3435.977  &0.000& $-$0.670  & $-$0.98 & $-$1.10 & $-$0.78  & 1 \\
    3469.921  &0.514& $-$0.129  & $-$0.98 & $-$1.00 & $-$0.98  & 1 \\
    3539.587  &0.000& $-$0.760  & $-$0.98 & $-$0.98 & $-$0.98  & 15 \\
    3675.567 & 0.188 & $-$0.840 & $-$0.98 & $-$0.98 & $-$0.98  & 1 \\ 
    \midrule
    \multicolumn{7}{c}{{\ion{U}{II} (Z\,$=$\,92)}} \\
    \multicolumn{7}{c}{{log$\epsilon$(U) = A(U) = $-$1.92 }} \\
    \midrule
    3859.577 &0.036& $-$0.067 &  --- & --- & 1,4,5,6 \\

\hline
\footnotesize{$*$ Hyperfine structure }\\
\end{tabular}
\end{table}

\section{CS 31082-001 at the resolution of CUBES}

The CUBES instrument is designed to deliver a spectral resolving power of R\,$\gtrsim$\,20,000 in its high-resolution mode, yielding FWHM\,$\sim$\,0.136\,{\rm \AA}  with a minimum of 2.3-pixel sampling, in the near-IV region. Assuming a 2.3 to 2.7-pixel sampling in the design, this corresponds to half the sampling obtained with UVES of 5 pixels per resolution element. The goal for the wavelength coverage is from 3000\,{\rm \AA} up to 4050\,{\rm \AA}, with an overview of the instrument given by Cristiani et al. (2022).

CUBES will be installed at the Cassegrain focus of one of the Unit Telescopes of the VLT. 
The design features a two-channel spectrograph fed by an image slicer (with six slices) yielding two complementary
spectra from 3000-3520\,{\rm \AA} and 3460-4050\,{\rm \AA} in two CCD detectors. The estimated efficiency of CUBES should enable observations of stars with V$\sim$18.0, considerably expanding the sample of metal-poor stars observable in the near-UV.

Initial simulations of CUBES observations were presented by Ernandes et al. (2020) to explore the parameter space of the instrument design. These first simulations were applied to red giant and dwarf models with metallicities of [Fe/H]\,$=$\,$-$1.0 and $-$3.0. From their results the new elements reported here for CS~31082-001 (Ho, Yb) appear feasible with CUBES.

To illustrate the potential of CUBES for some of the other elements studied here, in Fig.~\ref{cubes} we show simulated CUBES observations of the \ion{Dy}{II} 3434.369~{\rm \AA} and \ion{Os}{I} 3267.945~{\rm \AA} lines with
$R$\,$=$\,22,000, binned and recomposed through the End-to-end simulator (upper panels);
the UVES spectra convolved to $R$\,$=$\,22,000 and with a half number of points corresponding to 2.5 pixels per 
resolution element (middle panels);  compared with the UVES data (lower panels), and demonstrating comparably good
constraints on the estimated abundances.

\begin{figure}
    \centering
    \includegraphics[width=3.3in]{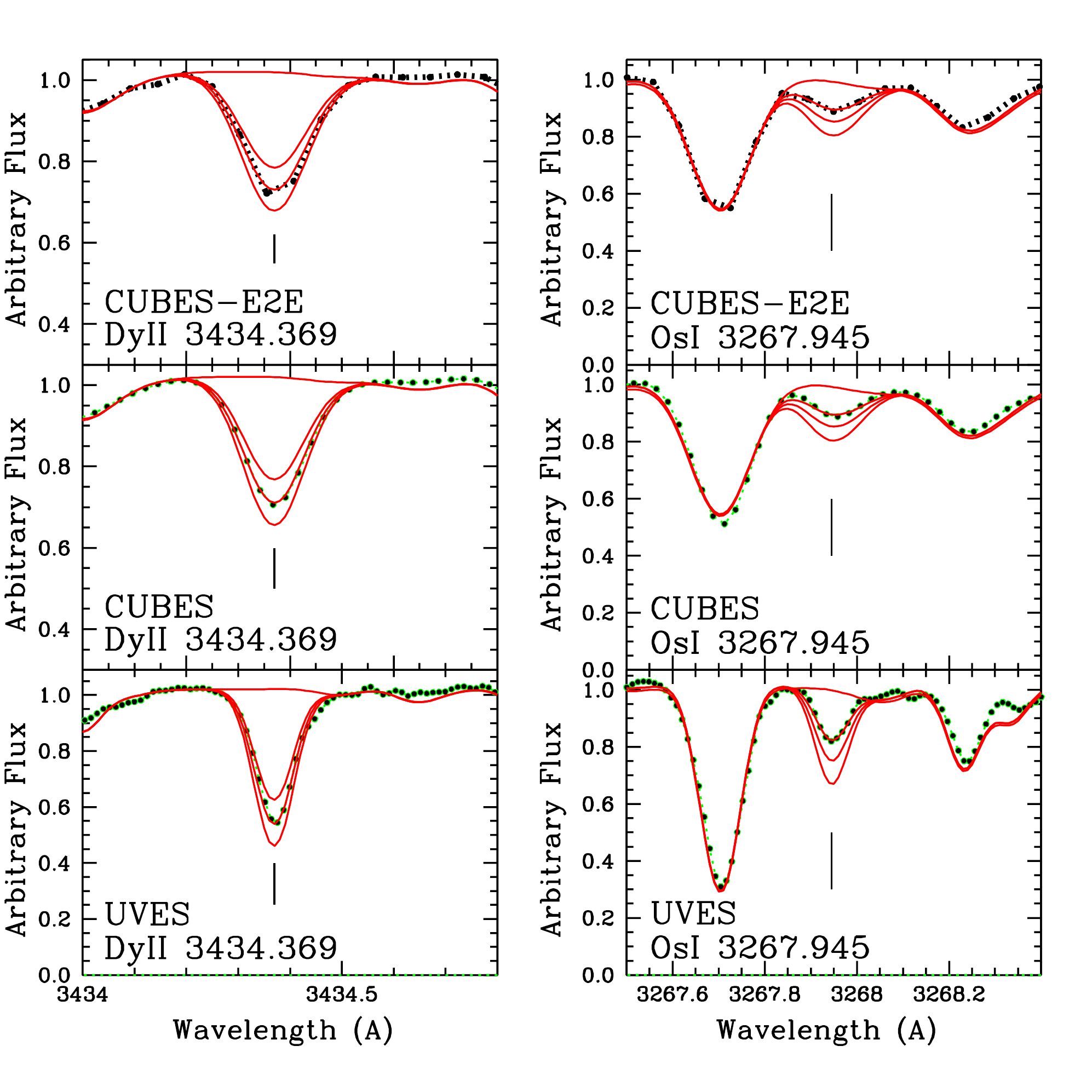}
    \caption{Simulated CUBES observations of CS~31082-001 for the \ion{Dy}{II} 3434.369 {\rm \AA} and \ion{Os}{I} 3267.945 {\rm \AA} lines binned and recomposed through the End-to-end simulator (upper panels); 
    $R$\,$=$\,22,000, 2.5 pixels per resolution element (middle panels), compared with the UVES data considered in this study (lower panels). The abundances are A(Dy)=none,$-$0.41,$-$0.21,$-$0.01 (left panels), and A(Os)=none,0.1,0.3,0.5 (right panels).}
    \label{cubes}
\end{figure}

\section{The MEAFS code}

We have gone through line-by-line fitting manually for the present results. In parallel, we are developing the
MEAFS code, created to automatically derive abundances with high precision for the lines that permit this treatment; 
a more detailed description will be given in a future publication. This optimised script is coded in Python with the key functions coded in C. The code generates synthetic spectra using a stellar spectral-synthesis software to automatically fit the best set of parameters using the Nelder-Mead (Singer \& Nelder, 2009) method to fit an observed spectrum. Our tests here of the MEAFS code for automatic, efficient, and precise measurements, have validated it for clear and unblended lines, as illustrated in  Fig.~\ref{LaTh}.

\begin{figure}
    \centering
    \includegraphics[width=3.3in]{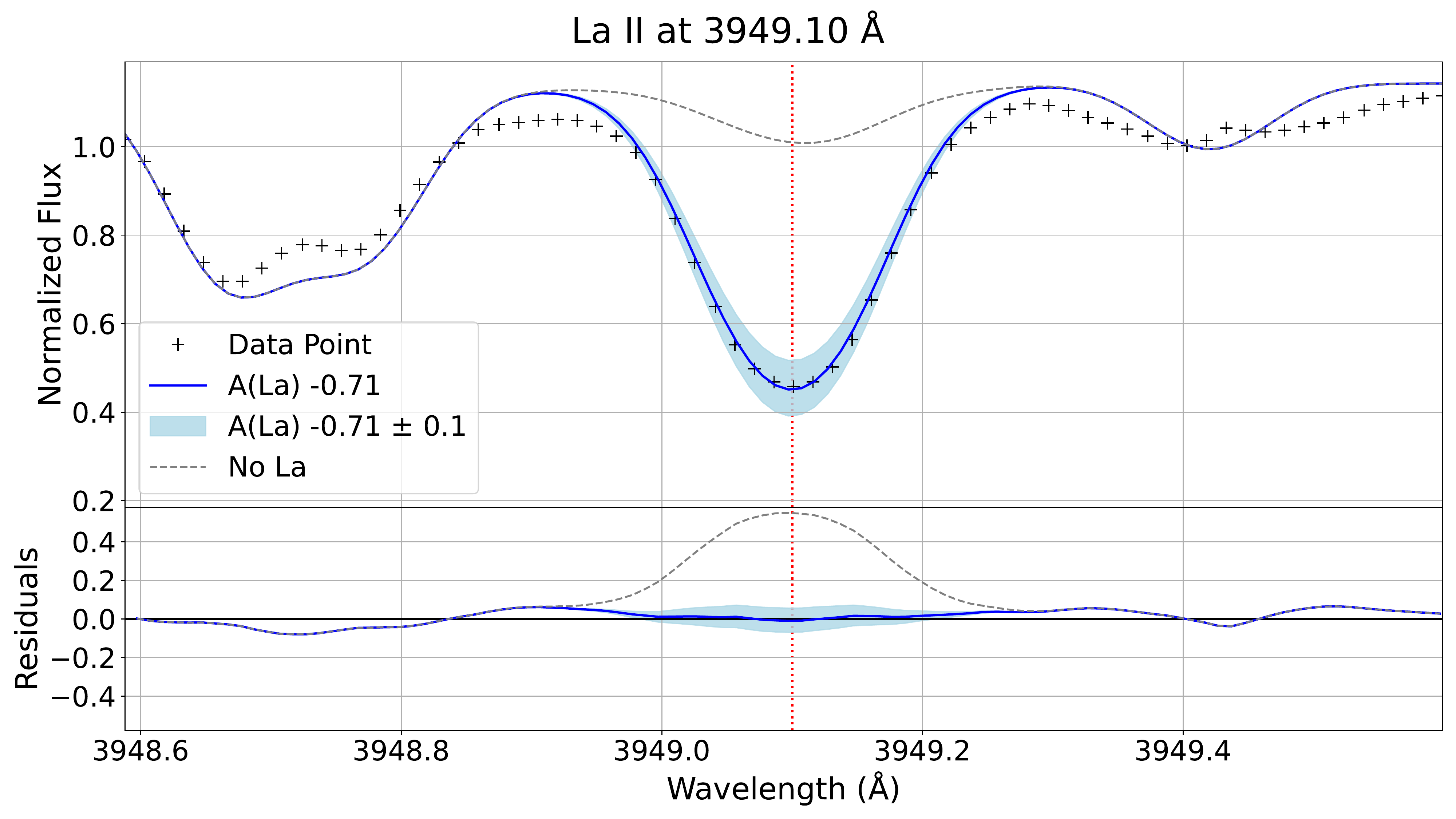}
    \includegraphics[width=3.3in]{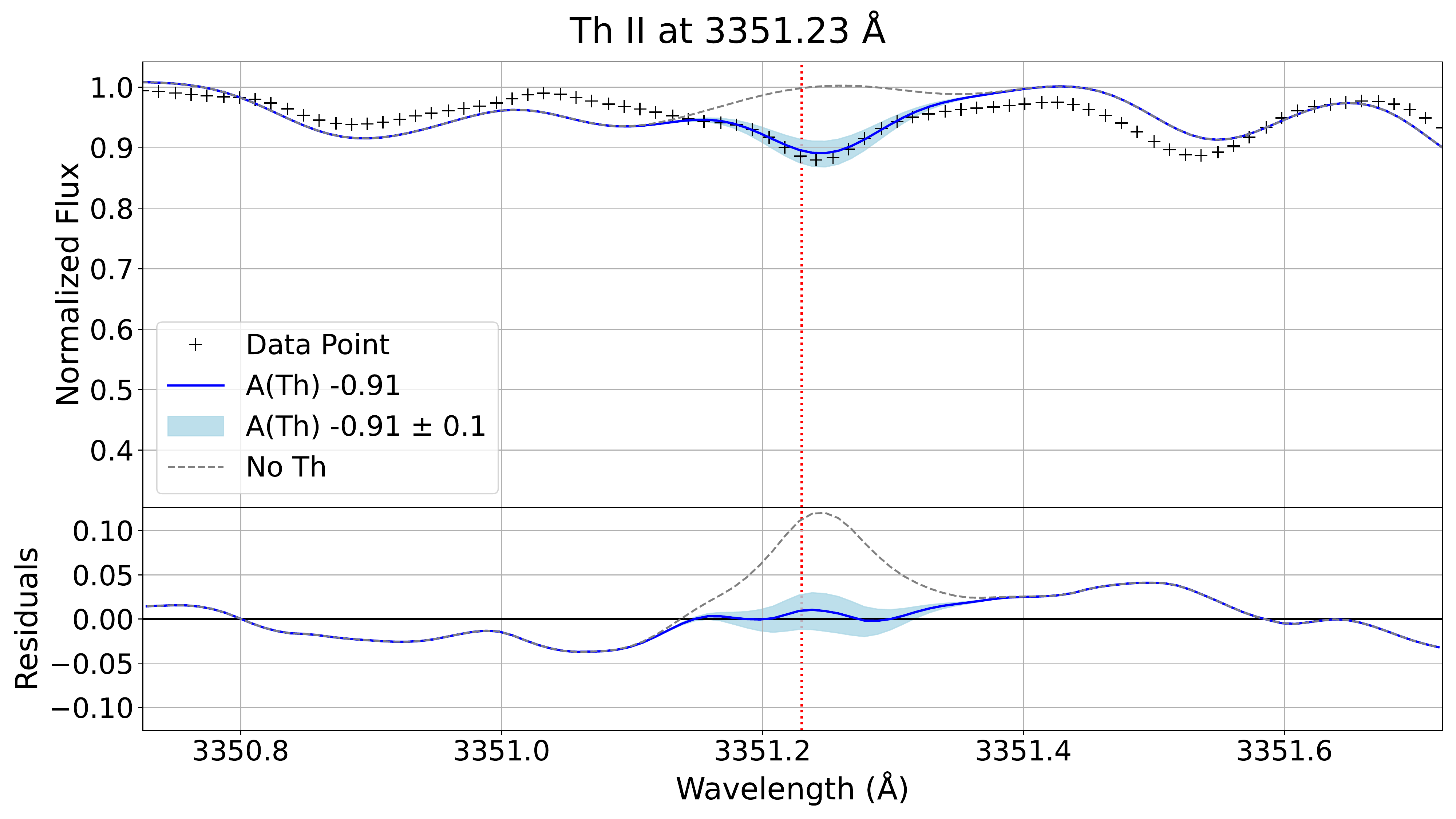}
    \caption{Fits of the \ion{La}{II} 3949.10~{\AA \rm} and \ion{Th}{II} 3351.23 {\AA \rm} lines using the MEAFS code.}\label{LaTh}
\end{figure}

%
\bsp	
\label{lastpage}
\end{document}